\documentclass[preprint,12pt, authoryear]{elsarticle}

\usepackage{graphicx,rotating,color}

\usepackage{amssymb}
\usepackage{amsmath}
\usepackage{float}
\usepackage{array}

\begin{document}

\begin{frontmatter}



\title{Networked relationships in the e-MID Interbank market: A trading model with memory}


\author[a]{Giulia Iori}

\author[b,c]{Rosario N. Mantegna\corref{*}}
\cortext[*]{Corresponding author: e-mail:
{\texttt{rn.mantegna@gmail.com}}}

\author[c]{Luca Marotta}

\author[c]{Salvatore Miccich\`e}

\author[a]{James Porter}

\author[d]{Michele Tumminello}

\address[a]{Department of Economics, City University London, London, United Kingdom}

\address[b]{Center for Network Science and Department of Economics,
Central European University, N\'ador u. 9, 1051 Budapest, Hungary}

\address[c]{Dipartimento di Fisica e Chimica, Universit\`a degli Studi di
Palermo\\ Viale delle Scienze, Ed 18. 90128 Palermo, Italy}

\address[d]{Department of Statistical and Mathematical Sciences, University of Palermo, Palermo, Italy}

\begin{abstract}
Interbank markets are fundamental for bank liquidity management. In this paper, we introduce a model of interbank trading with memory. Our model reproduces features of preferential trading patterns in the e-MID market recently empirically observed through the method of statistically validated networks. The memory mechanism is used to introduce a proxy of trust in the model. The key idea is that a lender, having lent many times to a borrower in the past, is more likely to lend to that borrower again in the future than to other borrowers, with which the lender has never (or has infrequently) interacted. The core of the model depends on only one parameter representing the initial attractiveness of all the banks as borrowers. Model outcomes and real data are compared through a variety of measures that describe the structure and properties of trading networks, including number of statistically validated links, bidirectional links, and 3-motifs. Refinements of the pairing method are also proposed, in order to capture finite memory and reciprocity in the model. The model is implemented within the Mason framework in Java. %
\end{abstract}


\tnotetext[acknow]{The authors acknowledge support from the FP7 research project CRISIS ``Complexity Research Initiative for Systemic InstabilitieS''.  G.I. acknowledge support from the FP7 research project FOC ``Forecasting Financial Crises ''.}

\end{frontmatter}


\section{Introduction} \label{intro}

Well functioning interbank markets effectively channel liquidity from
institutions with surplus funds to those in need and thus play a key role in banks liquidity management and the transmission of monetary policy. Before  the 2007-2008 financial crisis, liquidity and credit risks were perceived as negligible in these markets.  Nonetheless the collapse of interbank lending has been a central feature of the subprime financial crisis. Liquidity hoarding and  trust evaporation  have been identified   as  two important determinants of the  interbank market   drying up during the crisis (\citet{Heider2009}; \cite{Acharya2013}). \cite{Haldane2009} has advocated that the interbank market freeze is a  manifestation of the
behaviour under stress of a complex, adaptive network, the complexity arising from the interconnectedness of  players via mutual exposures to each other, and the adaptation from the attempts of agents to optimise interdependent strategies in the 
presence of (Knightian) uncertainty. 
Several authors have since  called for the adoption of network analysis to fully understand the stability of the banking sector.  

 Recently some theoretical studies have considered the problem of network formation in a financial system (\cite{Babus2007}, \cite{Allen2008}) and also, from the perspective of network formation games (\cite{Jackson1996}, \cite{Dutta2005}, \cite{Bloch2007}, \cite{Goyal2007}).  The presence of a network underlying the bilateral credit interactions occurring, for example, in an interbank market  has a role in the setting of both linkages that insure against liquidity risk and linkages that can channel contagion risk. 
 
 The empirical network literature has aimed  at characterising  the observed  topology of the interbank market checking for regularities and stylised facts. Studies have looked at several  interbank markets around the world including the  Austrian   (\cite{Boss2008}),  the e-MID  (\cite{Iorietal2008}, \cite{IazzettaManna2009}), the FedFunds (\cite{BechAtalay2008}), the German  (\cite{CraigVonPeter2010}, \cite{Brauning2011}),  the  Mexican (\cite{MartinezJaramillo2012}),  and the  Indian (\cite{IyerPedro2011}).  \cite{BechAtalay2008} show that the network resulting from Fedfunds exchanges is sparse, exhibits small world phenomena and is disassortative. In addition reciprocity and centrality measures are  predictors of  interest rate loans. The analyses of tiering in the German interbank market in \cite{CraigVonPeter2010} identifies a core-periphery structure. \cite{MartinezJaramillo2012} analyses the Mexican banking network combining data from the payment systems and interbank exposures. 
Other studies have attempted to quantify more directly the impact  of the network structure on the propagation of contagion addressing  a number of complementary issues including: the relationship between the network structure of the interbank market and its resilience to different kind of shocks (\cite{GaiHaldaneKapadia2011}, \cite{IoriEtAl2006}, \cite{BattistonEtAl2012}, \cite{LenzuTedeschi2012},  \cite{Georg2013}, \cite{Ladley2013} );
 the effects of assets fire-sale  (\cite{NierEtAl2007});  roll-over risk and portfolio overlaps (\cite{AnandEtAl2012}, \cite{CaccioliCatanachFarmer2012});  impact of regulatory taxes  (\cite{Thurner2013}, \cite{Poledna2014},) feedback loops between the macroeconomy and the financial sector (\cite{GrilliEtAl2012}). 
 Two complementary approaches have been adopted. In one case the exposure network is taken as exogenously given, either calibrated to real market data or generated according to preset specifications.  Stress-test experiments are then performed   and their effects monitored (\cite{GaiKapadia2010}, \cite{Upper2011}, \cite{CaccioliCatanachFarmer2012}).  Alternatively, in agent-based models, behavioural rules  are assigned to economic agents whose strategies determine endogenously both  the interbank exposure networks and default  events (\cite{IoriEtAl2006},  \cite{Georg2013}, \cite{Ladley2013}).  

Our paper belongs to the agent-based model tradition  and attempts to identify  simple behavioural rules that can explain features observed in real markets. In particular we are interested in the e-MID market, the only electronic market  for interbank deposits in the euro area and the USA. In a centralized interbank market, such as the e-MID, banks publicly quote their offers to lend or borrow money at a given maturity. The quotes can be anonymous but before finalizing the loan contract the lender has the right to know the identity of the borrower and can refuse to finalize the transaction. With this apparent lack of search frictions it is therefore worth investigating what network structure is present.

 While early studies on the e-MID market  (\cite{Iorietal2008}) have revealed a fairly random network  at the  daily scale, a non-random structure has been uncovered for longer aggregation periods.    Monthly and quarterly aggregated data show that  since the 1990s a high degree of bank concentration occurred (\cite{IazzettaManna2009}), with fewer banks acting as global hubs for the whole network. The hubs tend to cluster together and a significant core-periphery structure has been observed (\cite{Lux1}).   Lasting interbank relationships, which remained stable throughout the financial crisis,  have been observed by  \cite{Affinito2011} and \cite{Temizsoy2014}\footnote{ In these papers  the strength  of lending relationships is measured by the concentration of lending/borrowing activity between banks. More precisely, for every lender and  borrower a  preference index is computed, equal to the ratio of total funds that a lender (borrower)  has lent  to (borrowed from ) a borrower (lender) during a given period,
 over the total amount of funds that  the lender (borrower)  has lent  in (borrowed from)  the interbank market during the same period. This measure nonetheless does not take into account the heterogeneity of the banking system and the fact that large banks may have no alternative than to trade with each other if they need to exchange large volumes.}.
A networked structure of the e-MID market was observed in \cite{Hatzopoulos2013} by using a methodology based on the detection of statistically validated networks (\cite{Tumminello2011}) that allows the researcher to control for bank heterogeneity. In the cited study the networked nature of the e-MID market was highlighted by selecting repeated credit interactions (specifically, overnight loan contracts) between pairs of banks that were statistically incompatible with a null hypothesis of random pairing of the loans, which took into account the transaction heterogeneity of the banks.  In other words, the underlying trading network of banks was assumed to be primarily driven by the heterogeneity of the banks whereas the networked nature of some bilateral relationships was associated with a dynamical over-expression of the number of bilateral transactions. 
 
  Overall empirical studies of the e-MID data  have identified important properties of the market but  have also shown that  the e-MID interbank network remained surprisingly stable  during the subprime crisis (\cite{Lux2}) with a  structural break only appearing after the Lehman default.  These findings were confirmed, at the intraday scale, by  \cite{Vinko2013} who have uncovered   regularities in the network growth process that did not change during crisis. This indirectly suggests that the underlying  mechanisms driving the link formation are stable over time.
         
In the present study we introduce a stochastic model of preferential formation of bilateral credit relationships in a centralized heterogeneous credit market. The market heterogeneity in number of credit transactions is assumed to be exogenously given and the networked nature of credit relationships is associated with the detection of over-expression of bilateral transactions with respect to a null hypothesis taking into account banks' heterogeneity. The numerical simulations of the model are calibrated with real data from transactions of the e-MID market.  The simulations of our model show the existence of a dynamic of the networked credit relationships that mimics the ones observed in real data. The analysis of simulated data is performed by separately considering the role of lender aggressors from the role of the borrower aggressors. The aggressor is the party proposing the setting of the loan contract.  After calibration, in both cases the number of networked relationships observed in simulations agree well with the number observed in real data, both for lender aggressors and borrower aggressors.

In addition to the detection of the mean number of networked relationships, we also analyze the local nature of the networked relationships by studying the triads or 3-motifs present in 
simulations and real data both for the original and statistically validated networks.
Triads or 3-motifs, i.e. isomorphic classes of subnetworks of 3 nodes, have been recently investigated in empirical studies of the Interbank trading networks of the Netherlands (\cite{Squartini2013}) and Italy (\cite{Bargigli2013}). We show that real data presents an over-expression of some specific 3-motifs in several of the investigated three-maintenance periods whereas the over-expression is almost absent or only weakly present for lender-aggressor and borrower-aggressor statistically validated networks respectively. In other words, the real data presents a structuring of the local subgraphs that was not seen or only weakly seen in simulations of the calibrated model.       

The article is organized as follows. In Section \ref{model} we present our trading model. In Section \ref{methods} we describe the method used to reveal preferential trading relationships between banks. In Section \ref{data} we describe the investigated database of transactions at e-MID interbank market. In Section \ref{sec:comparison} we compare the trading networks obtained from simulations and from real data. In Section \ref{finitemem} a refinement of the model is proposed, in order to introduce a finite memory. Finally, in Section \ref{conclusions} we draw our conclusions. 

\section{A trading model with reinforcing memory} \label{model}

We investigate the credit relationships of bank loans in the e-MID market by considering them as a network in which nodes are banks and a directional link is set from bank $i$ to bank $j$ if $i$ lent money to $j$, the weight of the link being the number of times in which that event occurred in the considered time period.  We also distinguish between lender-aggressor  and borrower-aggressor transactions, thus investigating in several cases two distinct types of networks.

Here we propose a model of trading among banks that incorporates trust. This model relies on the idea that if a bank $B_j$ has lent many times to a bank $B_i$ ($i,j=1,\cdots,n$, where $n$ is the number of banks) in the past then it's likely that it keeps doing so in the future, unless external conditions change. Our model attempts to incorporate the intrinsic heterogeneity of banks with respect to their willingness to lend or borrow in a given time period. Therefore we shall distinguish two characteristic time scales in the model. The first time scale, $T_M$, is the one in which such willingness to trade is defined, e.g. a maintenance period or a three-maintenance period. Each bank can act as both a lender and a borrower in time window $T_M$. To best account for the heterogeneity of banks in our model we rely upon real data, and set the number of transactions that each bank $B_i$ is going to do as a lender (borrower) in \emph{lender}-aggressor transactions, $B_i^{l,la}$ ($B_i^{b,la}$), and as a lender (borrower) in \emph{borrower}-aggressor transactions, $B_i^{l,ba}$ ($B_i^{b,ba}$) 
from our real-data series of transactions.

The second time is an event time defining the order at which transactions of the model sequentially occurs. It is therefore naturally described in terms of time steps.
So, a time step $t$ of a certain time window $T_M$ indicates that $t$ transactions already occurred in the time window $T_M$.  The model shall distinguish between borrower-aggressor and lender-aggressor transactions. At each time step $t$ of time window $T_M$, we consider the outcome of a binary random variable $x_{lb}$, in order to decide if the next transaction is a lender-aggressor or a borrower-aggressor transaction. Variable $x_{lb}$ takes value ``lender-aggressor'' with probability $p_l(t)$, which is the fraction of the total number of lender-aggressor transactions that remain to do in time window $T_M$ at time step $t$, and value ``borrower-aggressor'' otherwise.  Once the decision about the type (lender-aggressor or borrower-aggressor transaction) of next transaction is made, the model indicates how the counterparts of the 
transaction should be selected. We first focus on a lender-aggressor transaction. To mimic a lender-aggressor transaction we assume that an order to borrow is placed by a bank $B_i$ that is randomly selected with probability 
\begin{equation}
\label{eq:la_borr_sel}
p_b(B_i,t) \propto B_i^{b,la}(t).
\end{equation}
Quantity $B_i^{b,la}(t)$ is the number of transactions that bank $B_i$ shall do in $T_M$ as a borrower in lender-aggressor transactions, $B_i^{b,la}$, minus the number of lender-aggressor transactions where it already acted as a borrower at time step $t$ of $T_M$. Once the borrower is selected, we randomly select a bank $B_j$ to be the counterpart of the transaction, the lender, with probability 
\begin{equation}
\label{eq:la_lend_sel}
p_l(B_j,t|B_i) \propto B_j^{l,la}(t)\,\left[w+N_{B_j \rightarrow B_i}(t)\right],
\end{equation}
where $B_j^{l,la}(t)$ is the number of transactions that a bank $B_j$ is willing to do in $T_M$ as a lender in lender-aggressor transactions, $B_j^{l,la}$, minus the number of lender-aggressor transactions in which it already acted as a lender at time step $t$. Quantity $N_{B_j \rightarrow B_i}(t)$ is the total number of transactions in which $B_j$ lent money to $B_i$ since the beginning of the simulations, in spite of the type of transaction, eventually including transactions from $B_j$ to $B_i$ among the $t$ trades that already occurred at time step $t$ of $T_M$. Finally, $w$ is a parameter equal for all the banks, which represents a common level of attractiveness of borrowers. This parameter is particularly important at the beginning of a simulation, when $N_{B_j \rightarrow B_i}(t)$ is equal to zero or it is very small. It is important to note two things about $w$. Firstly, it acts as a randomization parameter. In fact a large value of $w$ would prevent the memory mechanism from working effectively, and the result will 
be a random network without preferential links. Secondly, we may use parameter $w$ to include reputation in our model, by varying it across banks, and over time; however, this possibility is out of the scope of the present paper and will be explored elsewhere. The two equations above indicate that, according to our model, the probability that a (lender-aggressor) transaction occurs from $A$ (the lender) to $B$ (the borrower) at time step $t$ is
\begin{eqnarray}
p(B_j \rightarrow B_i,t) & = & p_b(B_i,t)\,p_l(B_j,t|B_i)\,\nonumber \\
  & = &\frac{B_i^{b,la}(t)}{\sum_{k=1}^{n} B_k^{b,la}(t)}\cdot \frac{B_j^{l,la}(t)\,\left[w+N_{B_j \rightarrow B_i}(t)\right]}{\sum_{q=1}^n B_q^{l,la}(t)\,\left[w+N_{B_q \rightarrow B_i}(t)\right]}.
\end{eqnarray}

The way in which we model borrower-aggressor transactions is slightly different, because the lender has always the possibility to refuse to trade with a specific borrower. The model works as follows. A lender, $B_j$, is randomly selected with probability
\begin{equation}
\label{eq:ba_lend_sel}
p_l(B_j,t) \propto B_j^{l,ba}(t),
\end{equation}
that is, with probability proportional to its willingness to lend in a borrower-aggressor transaction at time step $t$ of time window $T_M$. Then a borrower, $B_i$, is selected with probability proportional to its willingness to borrow as the aggressor at time step $t$ of time window $T_M$, that is $p_b(B_i,t) \propto B_i^{b,ba}(t)$. So far, the selection of lender and borrower occurred independently. However, once the borrower is selected, the lender, $B_j$ has a certain probability to accept borrower $B_i$ as counterpart in the transaction. This probability is set to be proportional to the attractiveness $w$ plus the degree of trust that lender $B_j$ associates with bank $B_i$. Therefore the probability that $B_j$ lends to $B_i$ at time step $t$ of time window $T$, conditioned to the fact that $B_j$ has been selected as the lender is:
\begin{equation}
p_b(B_i,t|B_j) \propto B_i^{b,ba}(t)\,\left[w+N_{B_j \rightarrow B_i}(t)\right].
\end{equation}
If the two banks do not end up trading then another borrower should be randomly selected, and so on, until the lender $B_j$ finds an acceptable counterpart. In our simulations, we work in the space of transactions\footnote{In fact time is just an event time increasing as an integer variable describing successive transactions}, and, from the perspective of transactions, the process of searching a suitable counterpart of lender $B_j$ is equivalent to randomly selecting a borrower $B_i$ with probability  
\begin{equation}
\label{eq:ba_borr_sel}
p_b(B_i,t|B_j)=\frac{B_i^{b,ba}(t)\,\left[w+N_{B_j \rightarrow B_i}(t)\right]}{\sum_{m=1}^n B_m^{b,ba}(t)\,\left[w+N_{B_j \rightarrow B_m}(t)\right]}.
\end{equation}
Therefore, in the case of borrower-aggressor transaction, we obtain that the probability that a transaction occurs between a lender $B_j$ and a borrower $B_i$ is given by:
\begin{eqnarray}
p(B_j \rightarrow B_i,t) & = &p_l(B_j,t)\, p_b(B_i,t|B_j) \nonumber \\
&=& \frac{B_j^{l,ba}(t)}{\sum_{q=1}^n B_q^{l,ba}(t)}\cdot \frac{B_i^{b,ba}(t)\,\left[w+N_{B_j \rightarrow B_i}(t)\right]}{\sum_{m=1}^n B_m^{b,ba}(t)\,\left[w+N_{B_j \rightarrow B_m}(t)\right]}.
\end{eqnarray}
The memory mechanism that we introduced in the model is based on the number of transactions between banks in the past. However, such a mechanism could be easily be adapted to take into account volumes: it is sufficient to replace,  in all the equations above, $N_{B_j \rightarrow B_i}(t)$ with the volume $V_{B_j \rightarrow B_i}(t)$ that bank $B_j$ lent to bank $B_i$ in the past.

\section{Statistically validated networks} \label{methods}

We asses the statistical significance of the observed interbank credit relationships by comparing the empirically observed number of transactions between each pair of banks against a random null hypothesis taking into account the trading heterogeneity of the system. As discussed in~\cite{Hatzopoulos2013}, we consider as null hypothesis a probabilistic description based on the hypergeometric distribution. It provides analytical results that are only approximated because the probabilistic description does not avoid the possibility that a bank can lend money to itself.  

For each link in the network, we perform a statistical test to check whether two banks preferentially traded in a given three-maintenance period. Our test is done by using a recently proposed method (\cite{Li2014}) that is a directional variant of the method presented in~\cite{Tumminello2011}. The statistical test is implemented as follows. For each three-maintenance period, we define $N_T$ as the total number of trades among banks in the system and focus on two banks $i$ and $j$ to check whether $i$ preferentially lent money to $j$. Let us call $n_{l}^i$ the number of times bank $i$ lent money to any other bank, and $n_{b}^j$ the number of times bank $j$ borrowed money from any other bank. Assuming that $n_{lb}^{ij}$ is the number of times bank $i$ lent money to $j$ then the probability of observing such $n_{lb}^{ij}$ trades, assuming that $j$ borrows money randomly and $i$ lends money randomly, is given by the hypergeometric distribution
\begin{eqnarray}
                               H(n_{lb}^{ij}|N_T,n_{l}^i,n_{b}^j)=\frac{{n_l^i \choose n_{lb}^{ij}} \, {N_T-n_l^i \choose n_b^j-n_{lb}^{ij}}}{{N_T \choose n_b^j}}.
\end{eqnarray}
We use this probability to associate a $p$-value with the observed number $n_{lb}^{ij}$ of trades from bank $i$ to bank $j$ as $p(n_{lb}^{ij}) =\sum_{X=n_{lb}^{ij}}^{\rm{min}[n_{l}^i,n_{b}^j]} H(X|N_T,n_{l}^i,n_{b}^j)$, that is the probability of observing by chance  a number of trades from $i$ to $j$ equal to $n_{lb}^{ij}$ or larger. This $p$-value is calculated by taking the sum of probabilities over the right tail of the hypergeometric distribution. Therefore a ``small"\footnote{This point will be quantitatively discussed later in this section.} value of the $p$-value statistically indicates that the link from $i$ (the lender) to $j$ (the borrower) is over-represented, in terms of number of loans, with respect to the null hypothesis of random trading. Analogously we can  statistically validate a link between $i$ and $j$ that is under-represented with respect to a random null hypothesis by taking into account the left tail of the hypergeometric distribution. In this case one should compute the left-tail $p$-value as $p(
n_{lb}^{ij}) =\sum_{X=0}^{n_{lb}^{ij}} H(X|N_T,n_{l}^i,n_{b}^j)$.

The hypergeometric distribution can be used to describe variable $n_{lb}^{ij}$ because the problem can be mapped into an urn model  \cite{Feller1968}, \cite{Hatzopoulos2013}. Note that mapping the problem of randomizing bank loans onto an urn model is done at the cost of removing the 
constraint that a bank cannot lend money to itself. This means that our analytical solution for the random system is an approximation of what we would obtain by randomly rewiring data and enforcing the condition of no self loans.

If we were just interested in calculating $p$-values of over-representation for all the directed edges, $E$ in our network, then we should run $E$ statistical tests. 
To avoid a large number of false positive validated links, due to the large number $E$ of statistical tests, it is advisable to consider a method to control the family-wise error rate. This control can be done by applying the so-called Bonferroni correction. This correction requires that the univariate level of statistical significance, e.g. $p_u=0.01$, is corrected in presence of multiple tests.

\subsection{Simultaneous test of over- and under-expression}
 
Here we investigate simultaneously over-expressed and under-expressed links of a network. This clearly affects the way in which the statistical threshold should be corrected for multiple comparison, because it changes the total number of tests. Below we show a simple way to compute a statistical threshold $p_m$ in this case.

Let us consider a specific time period --- e.g. a three-maintenance period $a$ --- and call $E_a$ the number of pairs of banks that actually traded at least once in that time period; call $L_a$ ($B_a$) the number of banks that have done at least one trade as a lender (borrower) in the time window, and  call $N_a^{BL}$ the number of banks that have done at least one trade as a lender and at least one trade as a borrower in the $a$-th time period. 

The number of tests for over-expression that we need to run is $E_a$. In general $E_a \le L_a\times B_a-N_a^{BL}$. In fact, the quantity $L_a\times B_a-N_a^{BL}$ gives the maximum expected number of links, given a certain number of lenders and borrowers. However, many of these links might not actually exist in the network, because trades between some bank pairs may not occur. Therefore, for some $i$ and $j$ it is possible that $n_{lb}^{ij}=0$, which, necessarily, cannot indicate an over representation of loans from $i$ to $j$, and therefore should not be tested for over representation. However, $n_{lb}^{ij}=0$ is an excellent candidate to test under representation of loans from $i$ to $j$---that is to test whether $i$ avoids to lend money to $j$ or $j$ avoids to borrow money from $i$. This means that such cases should be included in the test for under representation. So, in the case of under representation, we have to run the test for all the pairs of active banks in the system ($L_a\times B_a$) minus the 
cases in which lender and borrower are the same bank, that is $N_a^{BL}$. Therefore $L_a\times B_a-N_a^{BL}$ is  the number of tests of under-representation that we have to run. As a result, the total number of tests, for both under- and over-representation, in a given time window, is 
\begin{eqnarray}
                               T_a=E_{a}+L_a\times B_a-N_a^{BL}.
\end{eqnarray}
It is worth mentioning that a possible alternative would be $T_a=2 (L_a\times B_a-N_a^{BL})$. However we believe that this choice is unnecessariliy conservative as it involves testing the null hypothesis in cases that are clearly not relevant for the over-expression analysis (those with $n_{lb}^{ij}=0$).

In the present study, we investigate simultaneously over-expressed and under-expressed links. This is done to obtain results consistent with the results already presented in \cite{Hatzopoulos2013}. However, we will show only the results about over-expressed links. The investigation of under-expressed links is left for a future study.

\section{Data} \label{data}

Interbank markets can be organized in different ways: physically on the floor, by bilateral interactions,  or on electronic platforms. In Europe,  interbank trades are executed in all these ways. 
The only electronic market for Interbank Deposits in the euro area and USA is called e-MID. It was founded in Italy in 1990 for Italian Lira transactions and became denominated in Euros in 1999. When the financial crisis started, the market players were 246, belonging to 16 EU countries: Austria, Belgium, Switzerland, Germany, Denmark, Spain, France, United Kingdom, Greece, Ireland, Italy, Luxembourg, Netherlands, Norway, Poland, and Portugal. 
As shown in Fig \ref{fig1} and Fig. \ref{fig2}, the number of transactions at e-MID decreased, whereas the volume increased, until the beginning of the financial crisis. According to the European Central Bank e-MID accounted, before the crisis, for 17\% of total turnover in unsecured money markets in the Euro Area. The last report on money markets ~\citep{ECB},  recorded around 10\% of the total overnight turnovers.
Trading in e-MID starts at 8 a.m. and ends at 6 p.m. Contracts of different maturities,  from one day  to a year can be traded but the overnight segment  (defined as the trade for a transfer of funds to be effected on the day of the trade and to return on the subsequent business day at  $9:00$ a.m.) represents more than 90\% of the transactions. 
One distinctive feature of the platform is that it is fully transparent. Trades are public in terms of maturity, rate, volume, and time. Buy and sell proposals appear on the platform with the identity of the bank posting them (the quoter may choose to post a trade anonymously but this option is rarely used). Market participants can choose their counterparts.  An operator willing to trade can pick a quote and manifest his wish to close the trade while the quoter has the option to reject an aggression.
The database is composed by the records of all transactions registered in the period from 25-Jan-1999 to 7-Dec-2009. 
Each line contains a code labeling the quoting bank, i.e.\ the bank that proposes a transaction, and the aggressor bank, i.e.\ the bank that accepts a proposed transaction. The rate the lending bank will receive is expressed per year; the volume of the transaction is expressed in millions of Euros. The banks are reported together with a code representing their country and, for Italian banks, a label that encodes their size, as measured in terms of total assets. A label indicates the side of the aggressor bank, i.e.\ whether the latter is lending/selling (``Sell'') or borrowing/buying (``Buy'') capitals to or from the quoting bank. Other labels indicate the dates and the exact time of the transaction and the maturity of the contract. We consider the dataset obtained by considering only the overnight ({}``ON'') and the overnight long ({}``ONL'') contracts. The latter is the version of the ON when more than one night/day is present between two consecutive business day. This is the same dataset already 
investigated by \cite{Hatzopoulos2013}.  It is worth pointing out that in the present study and in \cite{Hatzopoulos2013} we limit our investigation to transactions occurring only between Italian banks.
%

The period of time in which credit institutions have to comply with the minimum reserve requirements is called the reserve maintenance period. During each reserve maintenance period minimum reserve levels are calculated on the basis of banks' own balance sheet. Each reserve maintenance period is usually equivalent to one calendar month, i.e. about 23 trading days. In the investigations we present below, we have aggregated the maintenance periods in groups of three. In fact, these aggregated periods better capture the natural economic cycles that are usually organized on a nearly 3-monthly basis. We therefore will consider 44 three-maintenance periods ranging from 25-Jan-1999 to 07-Dec-2009\footnote{It should be noted that the first three-maintenance period covers the time period from 25-Jan-1999 to 23-Mar-1999, thus involving a number of trading days which is reduced of a factor of about one third with respect to all the other three-maintenance periods.}.
%

\section{Comparison between model simulations and real data}  \label{sec:comparison}

\subsection{Original and Bonferroni networks} \label{sec:comparison1}

\begin{figure}
\begin{center}
                      {\includegraphics[width=10cm]{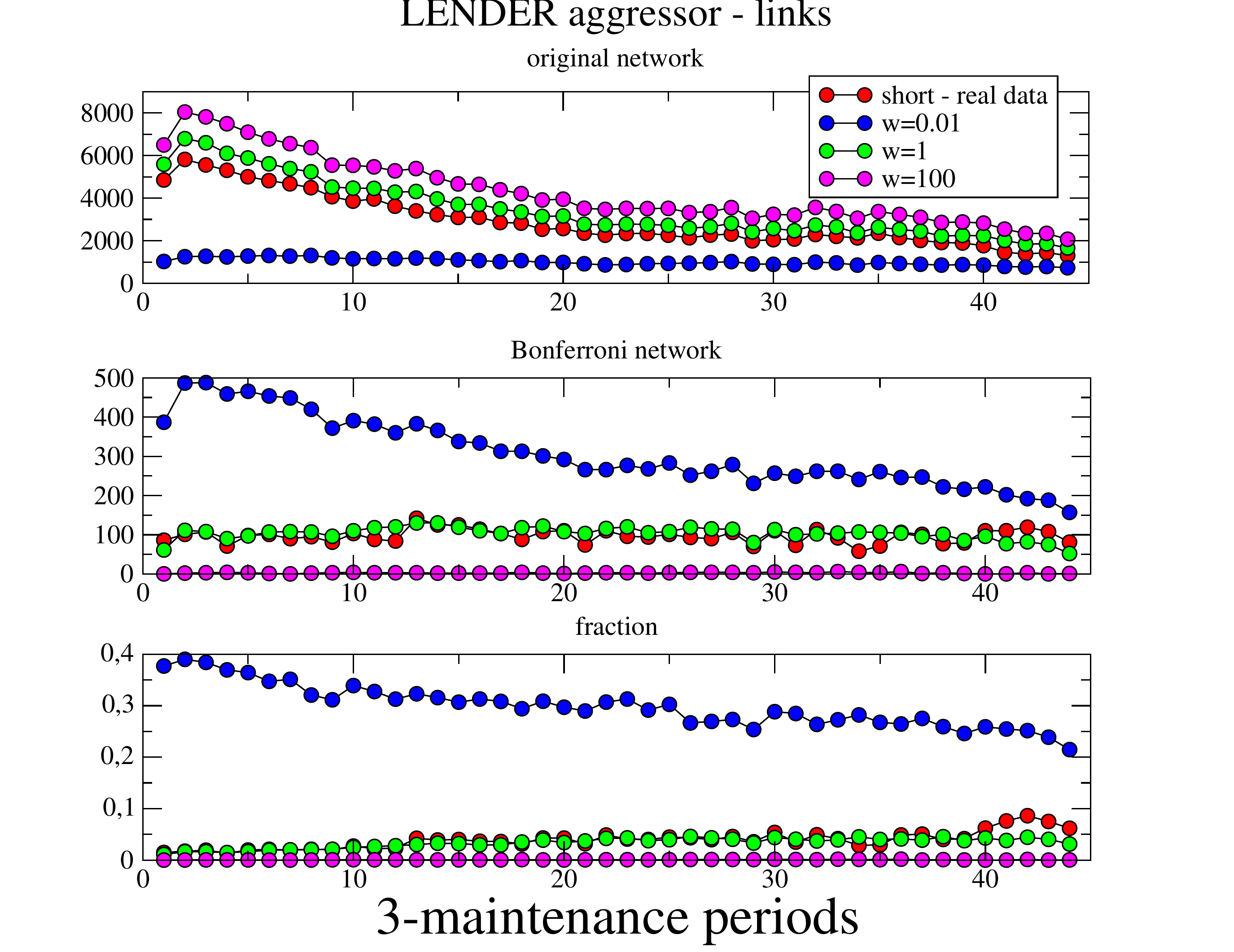}
                       \includegraphics[width=10cm]{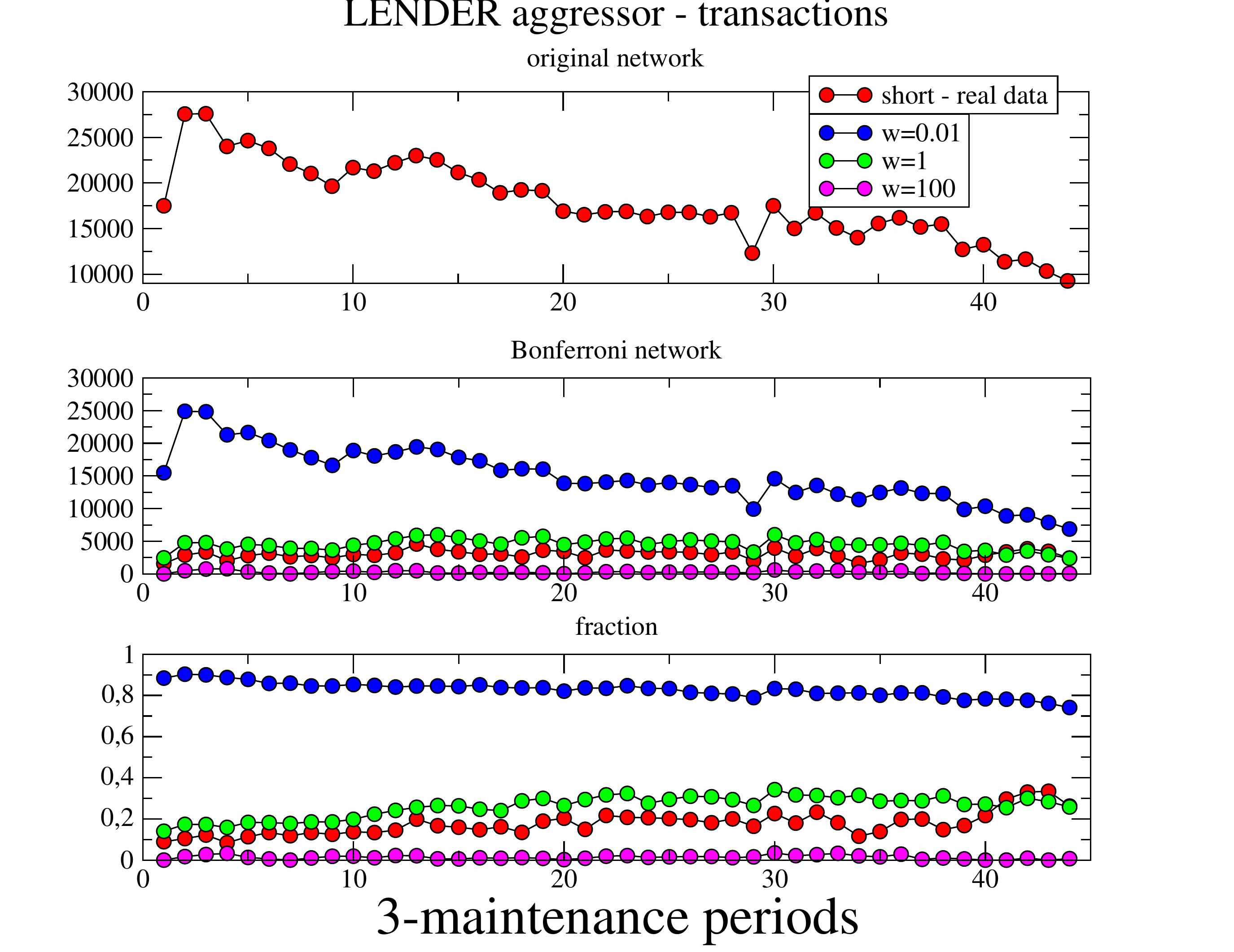}}
\caption{Top three panels: number of links in the original (top) and Bonferroni (central) network associated with lender-aggressor transactions simulated according to the presented model, for several values of parameter $w$. Specifically we have considered $w=0.01$ (blue circles), $w=1$ (green circles) and $w=100$ (magenta circles). The red circles refer to the empirical data. The bottom panel shows the ratio between number of links in the Bonferroni and original network. Bottom three panels: total number of transactions associated to the links of the original (top) and Bonferroni (central) network.  The bottom panel shows the ratio between number of transactions in the Bonferroni and original network.} 
\label{fig1} 
\end{center}
\end{figure}

\begin{figure}
\begin{center}
                      {\includegraphics[width=10cm]{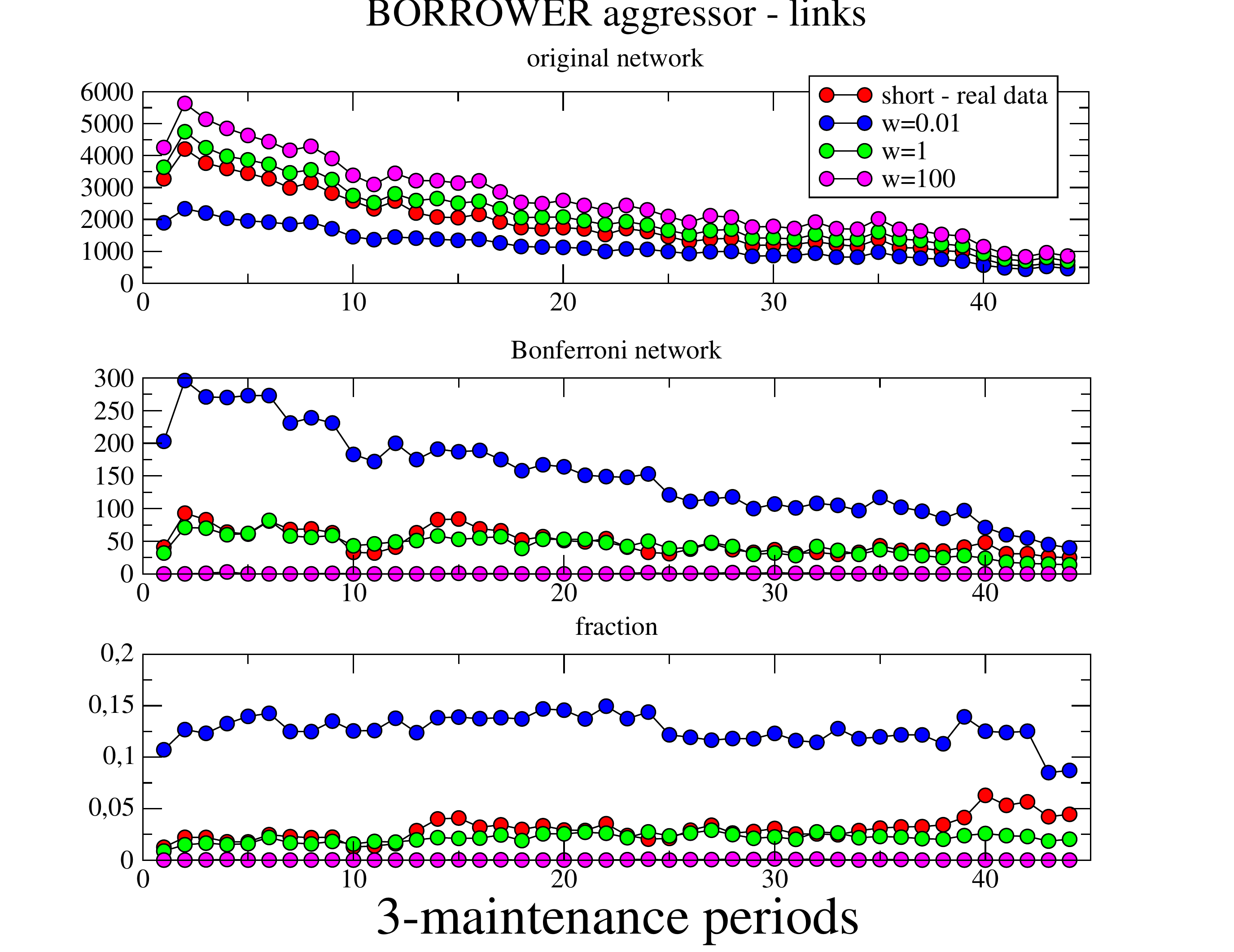}
                       \includegraphics[width=10cm]{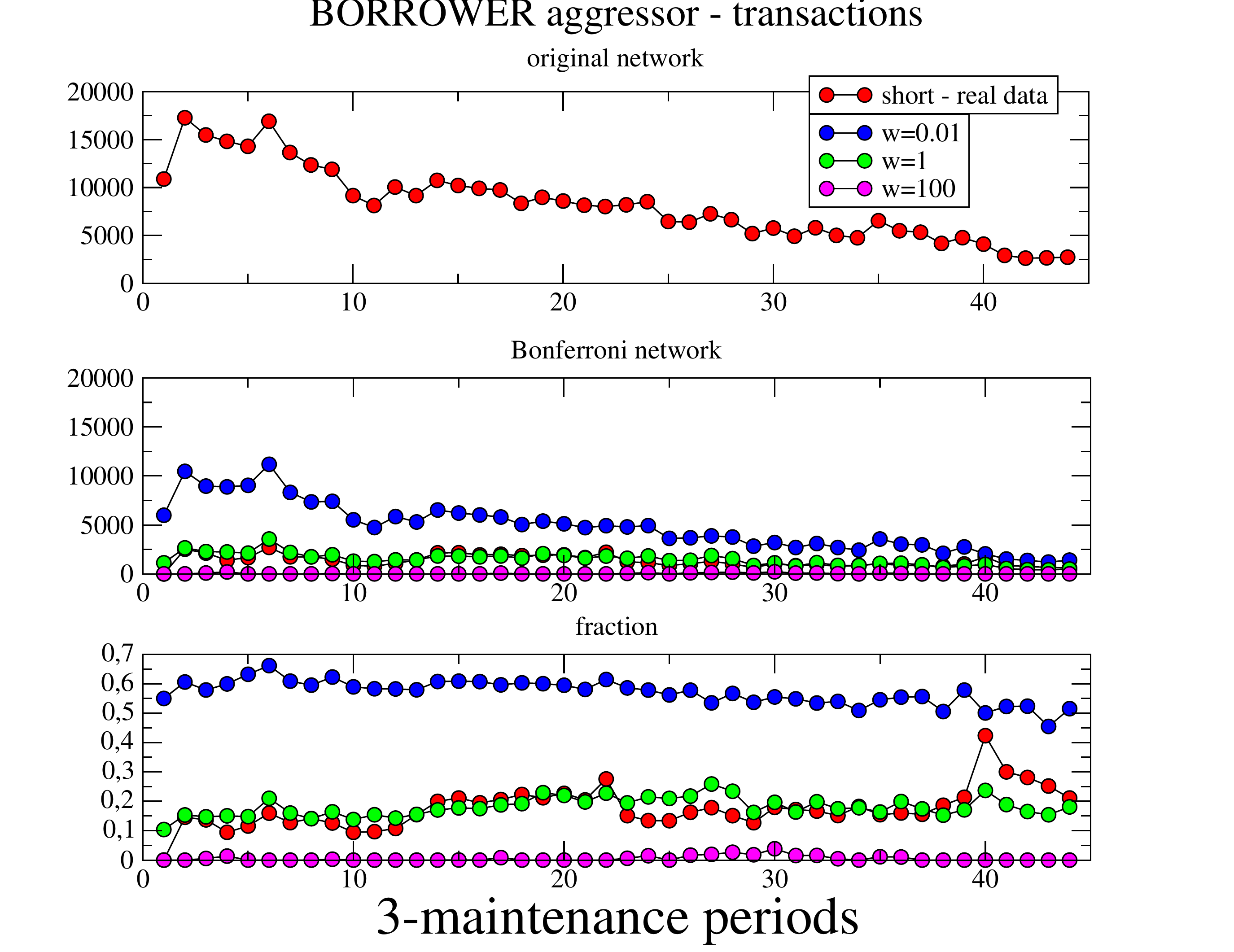}} 
\caption{Top three panels: number of links in the original (top) and Bonferroni (central) network associated with borrower-aggressor transactions simulated according to the presented model, for several values of parameter $w$. Specifically we have considered $w=0.01$ (blue circles), $w=1$ (green circles) and $w=100$ (magenta circles). The red circles refer to the empirical data. The bottom panel shows the ratio between number of links in the Bonferroni and original network. Bottom three panels: total number of transactions associated to the links of the original (top) and Bonferroni (central) network. The bottom panel shows the ratio between number of transactions in the Bonferroni and original network.}  \label{fig2} 
\end{center}
\end{figure}

In Figures \ref{fig1} and \ref{fig2}, the number of links in the original and Bonferroni networks of banks obtained by running and analyzing simulations of our model\footnote{Results presented in this section were obtained through an implementation of the model written by using Mathematica. The model has also been developed using Java and the MASON library for multi-agent modeling, and is being currently tested. 
 The implementation within the Java/Mason framework will allow to include and integrate our model into the interbank sector of the CRISIS macro-financial software library. A description of the implementation in Java/Mason is provided in Appendix A.}, for different values of parameter $w$, are compared with the number of links in the original and Bonferroni networks associated with real data of overnight transactions among Italian banks. Quantities $B_i^{l,la}$, $B_i^{b,la}$, $B_i^{l,ba}$, and $B_i^{b,ba}$ ($i=1,\cdots,n$), which are used in the model to set the initial willingness of bank $A$ to trade as a lender and as a borrower, in the two types of transactions---borrower-aggressor and lender-aggressor---are those observed in real data. The time window $T_M$ is equal to one three-maintenance period. The figures show that our model is capable to generate a networked market whenever parameter $w$ is small enough. Indeed middle panels of Figures \ref{fig1} and \ref{fig2} show that small values of parameter $w$ imply large numbers of validated links. This is in agreement 
with the interpretation of $w$ as a randomization parameter that, when large, can impair the effectiveness of the memory mechanism. 
Figures \ref{fig1} and \ref{fig2} show that the setting of the networked over-expressed connections is rapid involving one or two three-maintenance periods. Such an abrupt change in the number of validated links indicates that the memory mechanism becomes fully effective in a few time windows. The figures also show that, when $w$ is small, the number of links in the Bonferroni networks of the simulations strongly correlates with the total number of transactions.  This correlation is explained by noting that small values of $w$ allow the memory mechanism to determine and enhance even small deviations from random matching, and such small  deviations are better detected by the statistical validation method when the statistics, that is, the total number of transactions, is large. On the contrary, larger values of $w$ may easily destroy small deviations from random matching, and the detection of larger deviations is less affected by varying statistics.  

According to the number of validated links, we obtain a good agreement between simulations and real data if $w$ is set equal to $1$. Specifically, we compared the average (over the 44 three-maintenance periods) number of validated links from simulations and real data through a two-sample Student's t-test \footnote{The test assumes that the two sets have equal sample size and unequal variance, and the $p$-value is obtained through $10^6$ bootstrap replicas of data}. The result is a two-tailed $p$-value of 0.047 for the lender-aggressor data set and a $p$-value of 0.156 for the borrower-aggressor data set, indicating that the hypothesis that the two samples, from simulations and real data, have the same average cannot be rejected at $1\%$ significance level. It is important to note that the process of statistically validating links is done independently for lender-aggressor and borrower-aggressor transactions, in both real data and simulations. Therefore the fact that one value of parameter $w$, $w=1$, allows one to 
replicate, on average, the number of validated links in a three-maintenance period for both lender-aggressor and borrower-aggressor transactions is not a consequence of the validation procedure. The two datasets of borrower-aggressor and lender-aggressor transactions are analyzed independently one of the other. The results obtained from simulations with $w=1$ indicate that the tradeoff between memory and the overall attractiveness of banks might be independent of the type of transactions, lender-aggressor or borrower-aggressor, in which they are involved.  

\subsection{Degree of similarity of the networks}

\begin{figure} [H]
\begin{center}{
                        \includegraphics[scale=0.38]{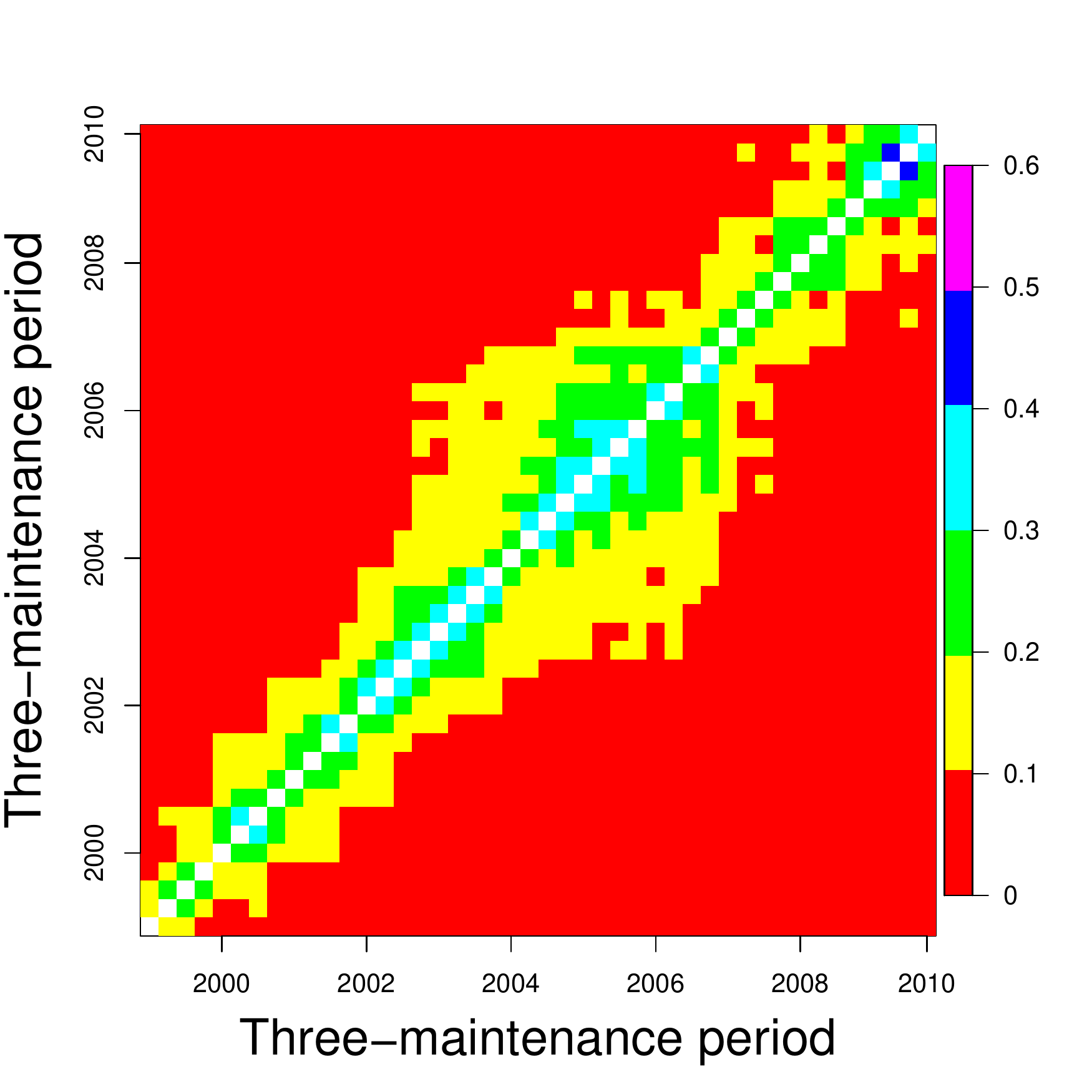} 
                        \includegraphics[scale=0.38]{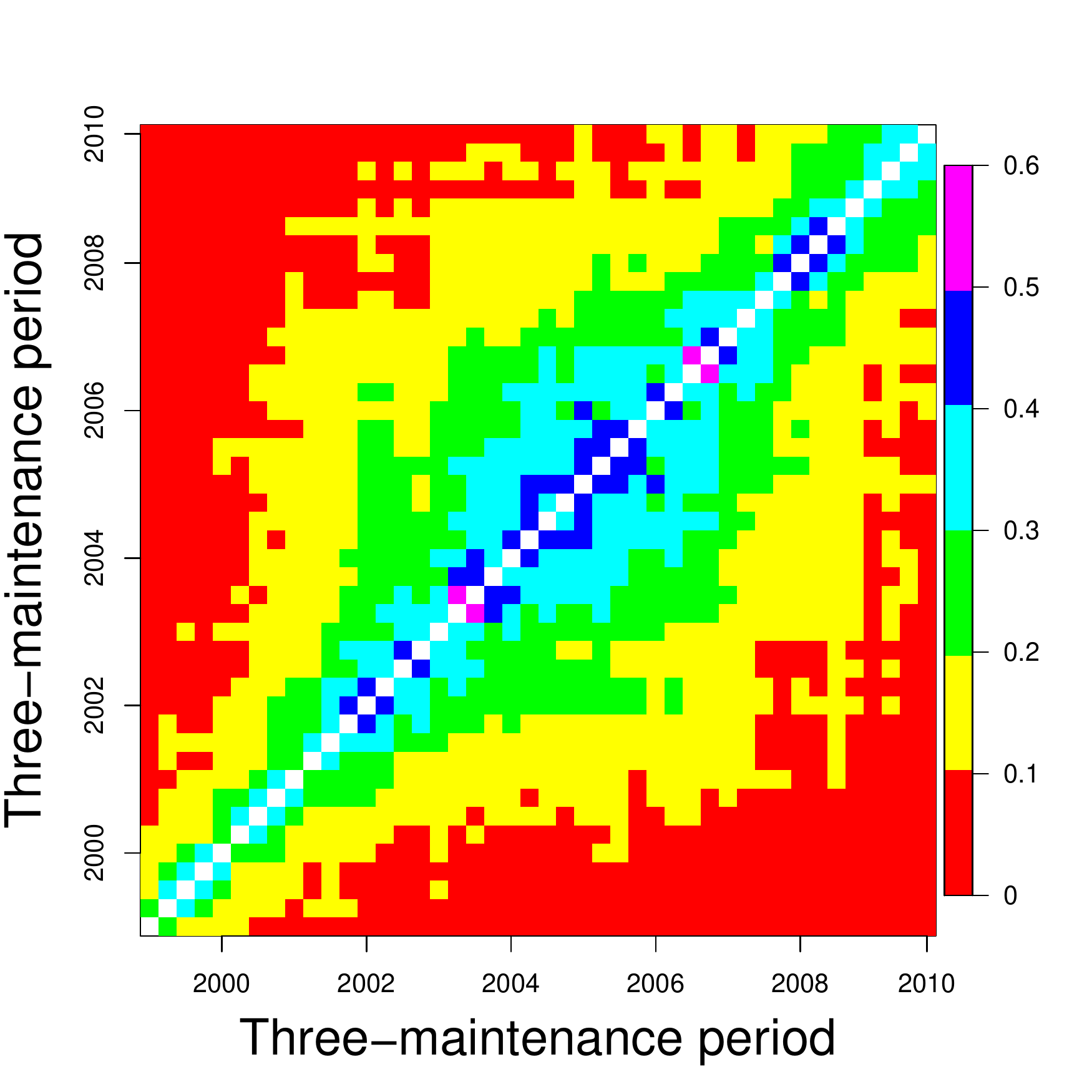}}\\
                         {
                         \includegraphics[scale=0.38]{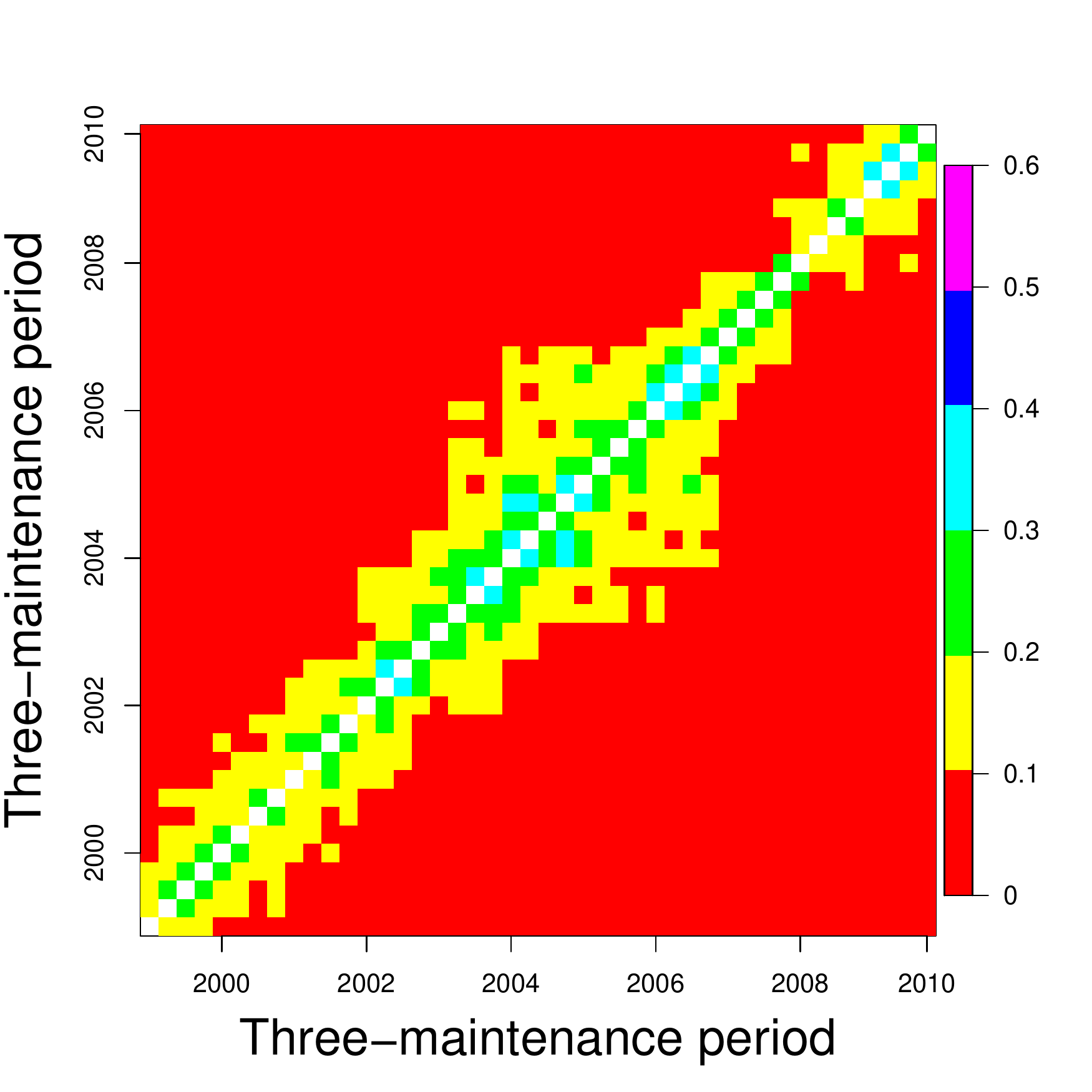} 
                         \includegraphics[scale=0.38]{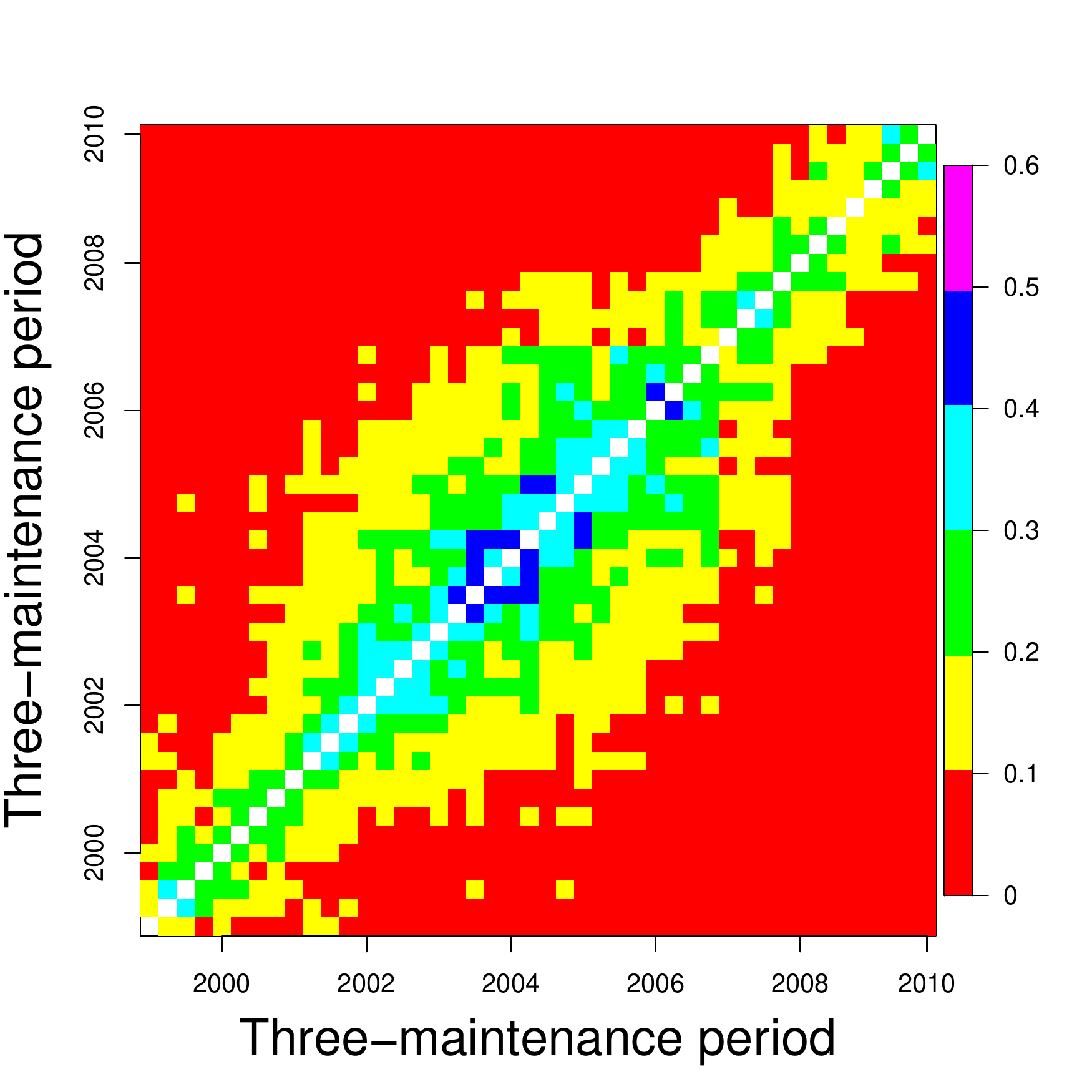}}
                         \caption{Matrix of Jaccard indices between validated networks associated with real data of overnight transactions among italian banks (left panels) and between validated networks associated with simulations of our model realized across 44 time windows with parameter $w=1.00$ (right panels). Lender-aggressor (borrower-aggressor) transactions are considered in the top panels (bottom panels).}  \label{jaccardMOD} 
\end{center}
\end{figure}

Another feature of the model is that the memory mechanism is cumulated, that is, the probability that two banks end up trading depends on the their trading history since the beginning of the simulation. In practice, this feature introduces a certain degree of similarity between transactions, and, therefore, between validated networks, obtained at different time windows $T_M$. We quantify this degree of similarity through the Jaccard index between any two validated networks, $g_1$ and $g_2$, obtained at different time windows:
\begin{equation}
J(g_1,g_2)=\frac{|E_1\cap E_2|}{|E_1\cup E_2|},
\end{equation}
where $|E_1\cap E_2|$ is the number of directed links that appear in both validated networks, and $|E_1\cup E_2|$ is the number of links that appear in either networks. In Fig.\ref{jaccardMOD}, we report the matrix of Jaccard indices between every pair of validated networks obtained from real data across the 44 three-maintenance periods (left panels), and the corresponding matrix obtained from simulations of the model with $w=1$ across the corresponding 44 time windows (right panels). Top (bottom) panels correspond to lender-aggressor (borrower-aggressor) transactions. The pattern of Jaccard index across the 44 three-maintenance periods analyzed for real data is similar to the one observed for simulations. Overall, the analysis of the Jaccard index indicates that validated networks obtained from simulations present a higher degree of similarity, i.e. of memory, than those obtained from real data. This evidence may suggest that the memory mechanism that we introduced in the model, which invoke the 
fact that banks keep memory of the whole set of their transactions in the past, may be suitably varied in such a way to require finite memory of banks. This can easily be done by redefining the quantity $N_{A \rightarrow B}(t)$, which is used to incorporate memory in the model, as the number of transactions in which bank $A$ lent money to bank $B$ in the past $Q$ time-windows. This refinement is considered Section  \ref{finitemem}. Another mechanism that may compete with finite memory to explain the differences between real data and simulations, which are observed in Fig.\ref{jaccardMOD}, is the nature of parameter $w$. This randomization parameter is maintained constant throughout the 44 time windows of a simulation. However, if we think about $w$ as a parameter that also incorporates market information, such as, for instance, market liquidity, it becomes reasonable to assume that $w$ can vary over time, in order to mimic the existence of different ``states'' of the market. 

\begin{figure}[H]
\centering
                {\includegraphics[scale=0.38]{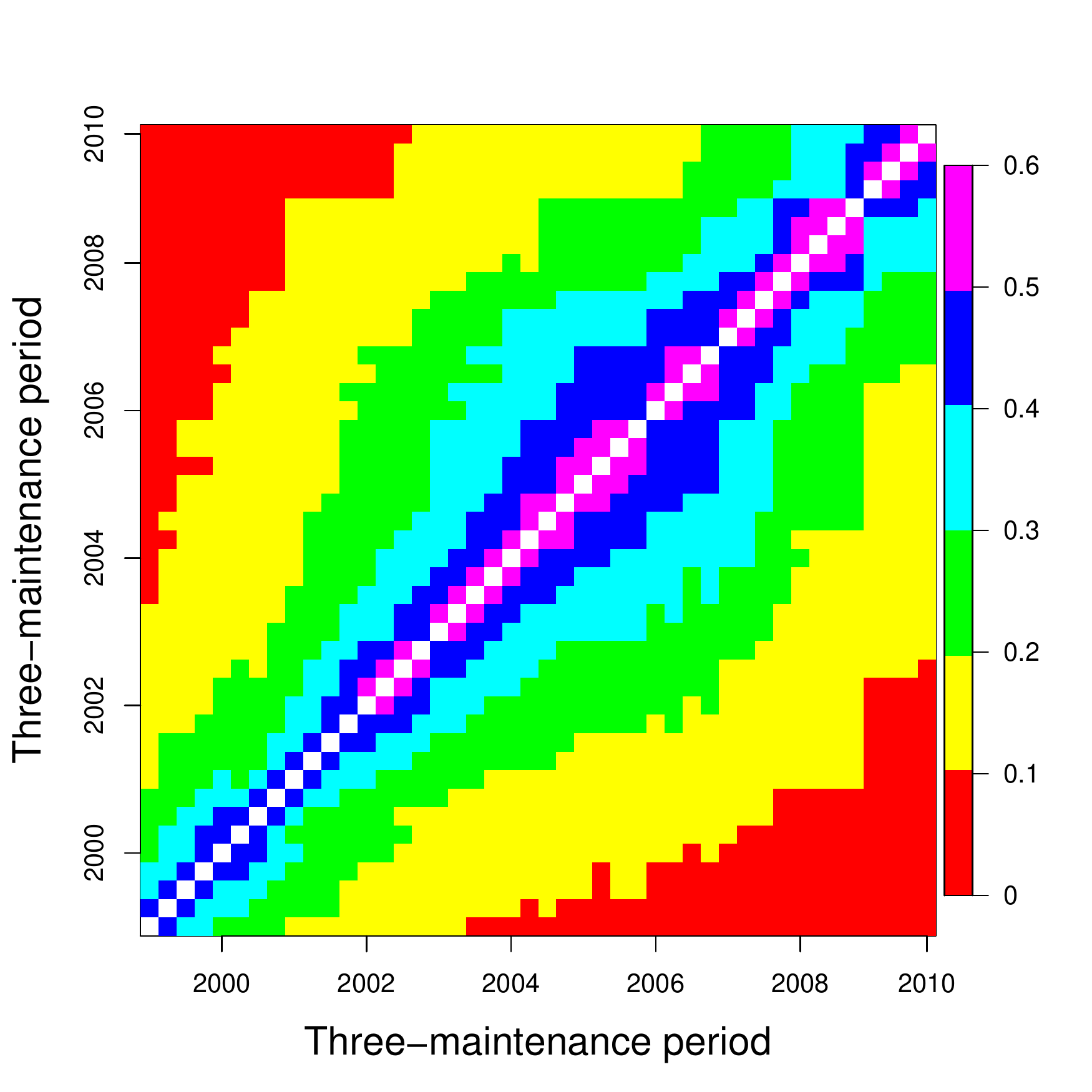}
                 \includegraphics[scale=0.38]{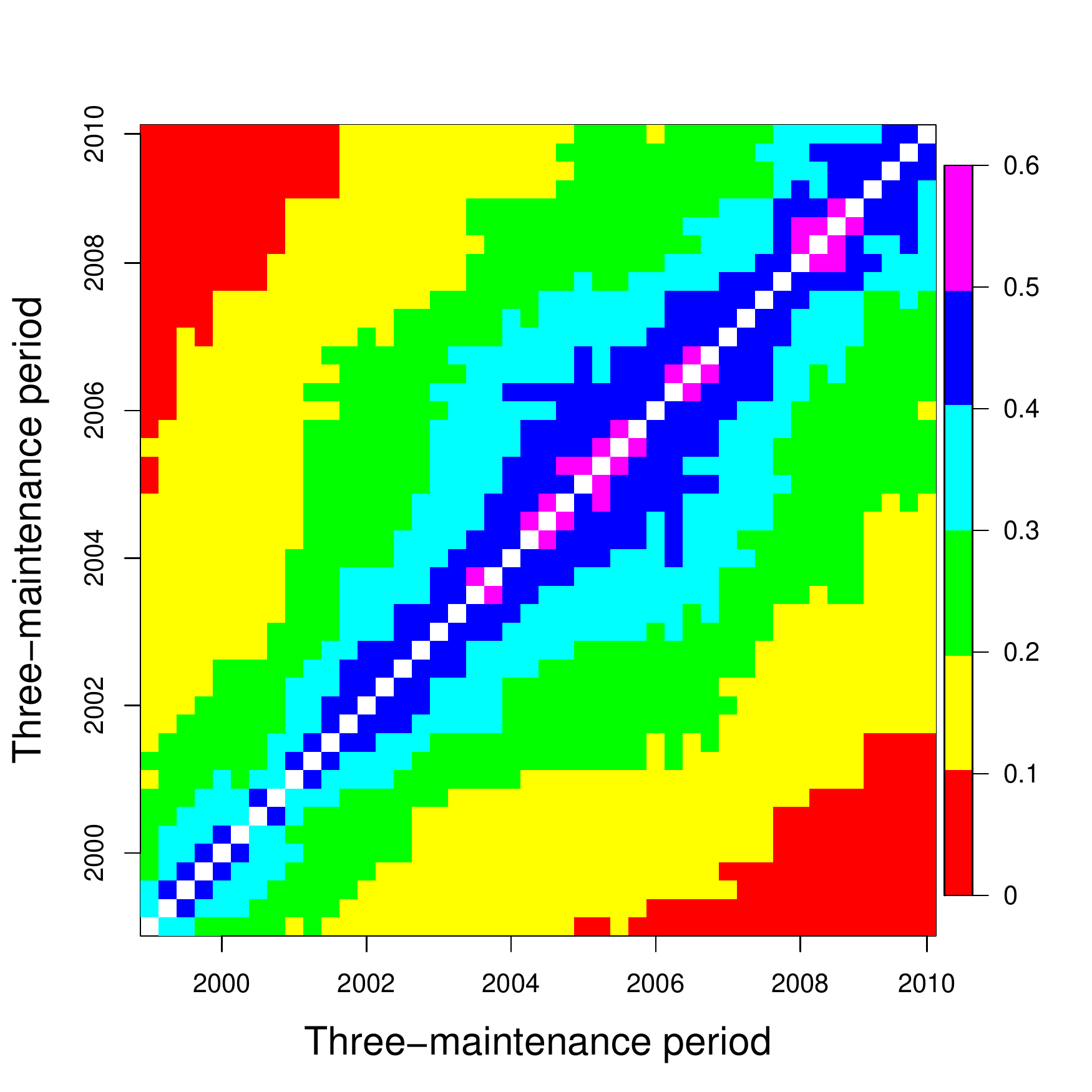}} 
                 \caption{Lender-aggressor networks. Matrix of Jaccard index of links between all pairs of networks of three-maintenance periods. The left panel shows the Jaccard index between original networks. The right panel shows the Jaccard index between the corresponding simulated networks with $w$=1.} \label{fig:M_contour}
\end{figure}

The differences between empirical and simulated networks can  also be investigated on the original and simulated networks over the 44 three-maintenance periods. In Fig. \ref{fig:M_contour} we show the the Jaccard index of lender-aggressor networks estimated between all pairs of three-maintenance periods. Specifically, the left panel shows the Jaccard index between original networks whereas the right panel shows the Jaccard index between simulated networks with $w$=1. Also at the level of the original network the simulations of the model well describe the degree of network persistence observed in real data.

\subsection{Bidirectional links}  \label{sec:comparison3}

Our model does not involve any mechanism of reciprocity. Indeed the memory term $N_{A \rightarrow B}(t)$ only counts the number of times in which bank $A$ lent to bank $B$ in the past, and, therefore, does not include information about the number of times in which $A$ borrowed money from $B$ in the past. This lack of reciprocity implies that a bidirectional link, either statistically validated or not, may appear in an outcome of the model only by chance. To check if our hypothesis of no reciprocity is consistent with real data, we compared the number of bidirectional links, in the original and Bonferoni network.
Table \ref{tabbid} shows that the average number of bidirectional links observed in real data is rather small, in both the original and Bonferroni network. Specifically the fraction of bidirectional links is always 
smaller than 8\% in real data. Such a small value justifies neglecting reciprocity in the basic setting of our model. However, our results show that the average number of bidirectional links in empirical data is always larger than the corresponding number of bidirectional links obtained from simulations with parameter $w=1$, in both lender-aggressor and borrower-aggressor datasets.

\begin{table}[ht]
\caption{Bidirectional links in real data and simulations with $w=1$}\label{tabbid}
\begin{center}
\begin{tabular}{|l|ccc|ccc|}
\cline{2-7}
\multicolumn{1}{c|}{} & \multicolumn{3}{c|}{Original network} & \multicolumn{3}{c|}{Bonferroni network}\\
  \hline
  Data type				& Mean &  Std. & Perc. & Mean &  Std. & Perc. 	\\
  \hline
Lender aggr. (data)	 	& 210.8 & 111.9 & 7.6\% & 1.64 & 1.59 & 1.7\% 		\\
Lender aggr. (sim.)	& 202.5 & 93.0 & 6.1\% &0.07  & 0.25 & 0.07\%		\\
Borrower aggr.	(data)	& 91.1 & 67.2 & 4.7\% & 0.45 & 0.85 & 1.2\%		 	\\ 
Borrower aggr.	(sim.)	& 88.5 & 61.8 & 3.8\% &0.00	& 0.00 & 0.0\%		\\
 \hline
\end{tabular}
\end{center}
\end{table}

The results shown in Table \ref{tabbid} suggest that it may be worth considering the possibility of introducing a reciprocity mechanism as a refinement of our model. This could be done by weighting the memory term $N_{A \rightarrow B}(t)$ with $N_{B \rightarrow A}(t)$:
$$
N^{\lambda}_{A \leftrightarrow B}(t) = \lambda\,N_{A \rightarrow B}(t)+(1-\lambda)\,N_{B \rightarrow A}(t),
$$ 
where $\lambda$ is a parameter ranging between 0 and 1. The quantity $N^{\lambda}_{A \leftrightarrow B}(t)$ can be used in place of $N_{A \rightarrow B}(t)$ in all the equations of the model in order to introduce a degree of reciprocity, which is controlled by parameter $\lambda$. Intuitively, the value of $\lambda$ should be quite close to 1, in order to replicate the (rather small) average number of bidirectional validated links observed in real data. 

\subsection{3-motifs}

We have also compared the original and statistically validated networks from real data and from simulations according to the fraction of 13 different types of 3-motifs that can be present in a network. 

\begin{figure}
\centering
                \includegraphics[scale=0.5]{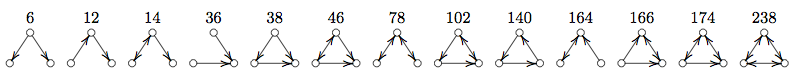}
                \caption{The 13 isomorphic directed 3-motifs. The numeric code is the one used by the FANMOD program.} \label{motifs}
\end{figure}

In Fig. \ref{motifs} we show the 13 different types of isomorphic 3-motifs. There are different ways to label 3-motifs. In the present paper we use the labeling of FANMOD program. 

\begin{figure} [H]
\begin{center}{
                       \includegraphics[scale=0.36]{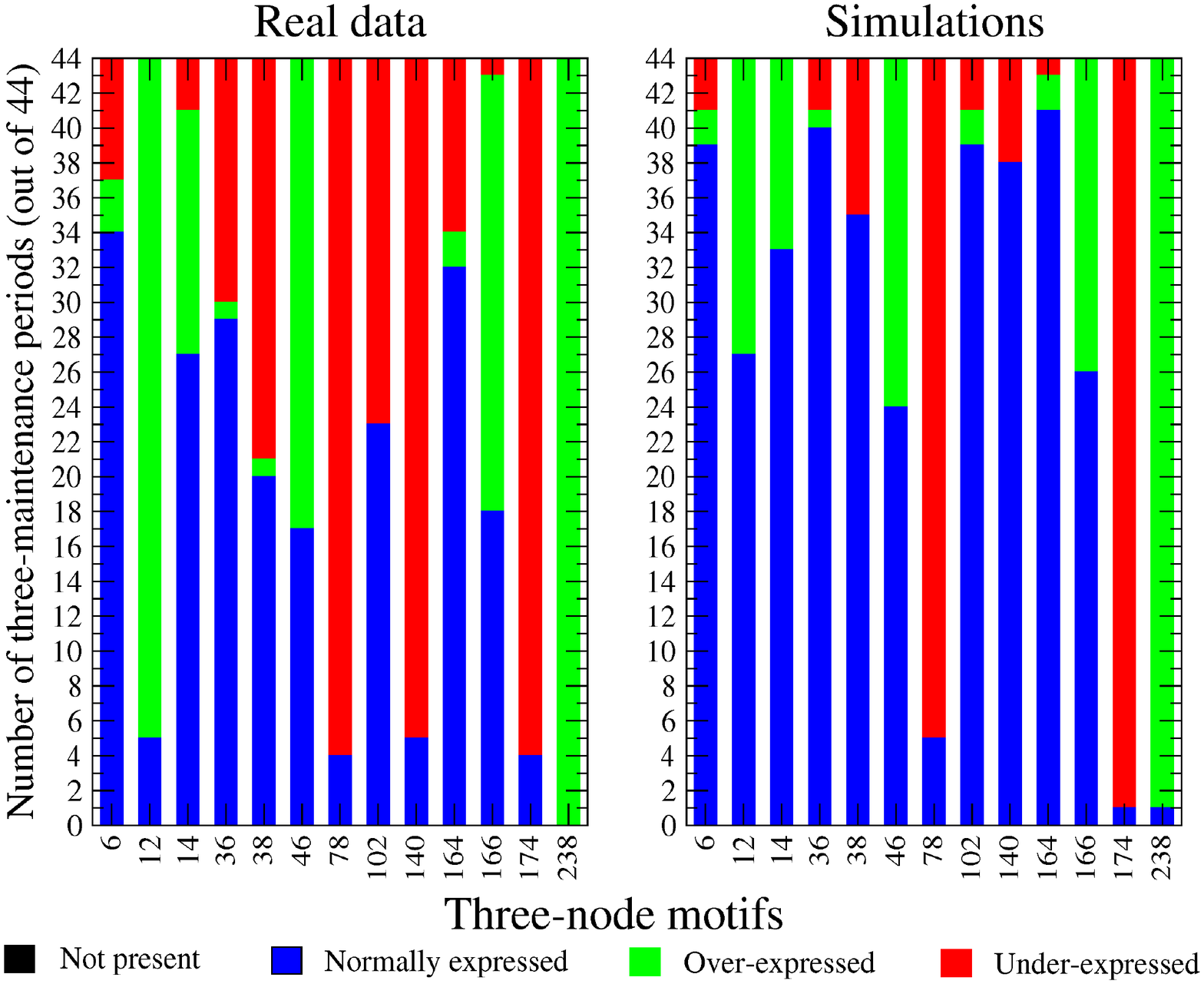}
\includegraphics[scale=0.36]{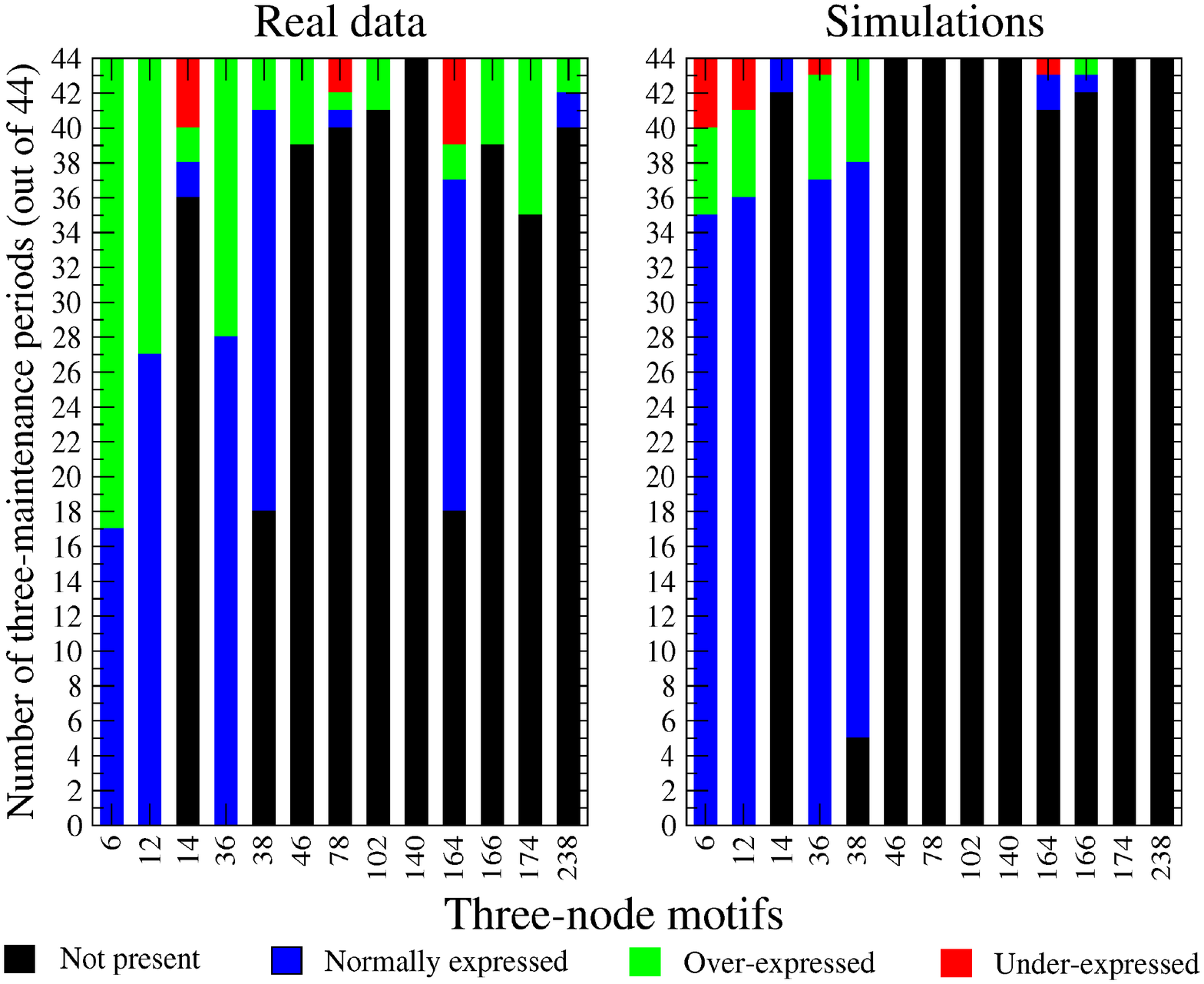}
}
                       \caption{Number of three-maintenance periods (out of 44) in which each three-motif type (indicated in the horizontal axis) is over-expressed, under-expressed, normally expressed, and not present in the original (top left panel) and Bonferroni networks (bottom left panel) associated with real data of \emph{lender-aggressor} and with corresponding \emph{lender-aggressor} transactions from simulations of the model with $w=1$ (original (top left) and Bonferroni (bottom left)). Over-expressions and under expressions are obtained by performing a multiple hypothesis test correction. The over/under expression of a three motif indicates that the corresponding $p$-value provided by FANMOD was smaller than $0.01/(13 \cdot 44)$, where $13$ is the number of three-motif types and $44$ is the number of three-maintenance periods investigated.}  \label{threemotifslend} 
\end{center}
\end{figure}

\begin{figure} [H]
\begin{center}{
                       \includegraphics[scale=0.36]{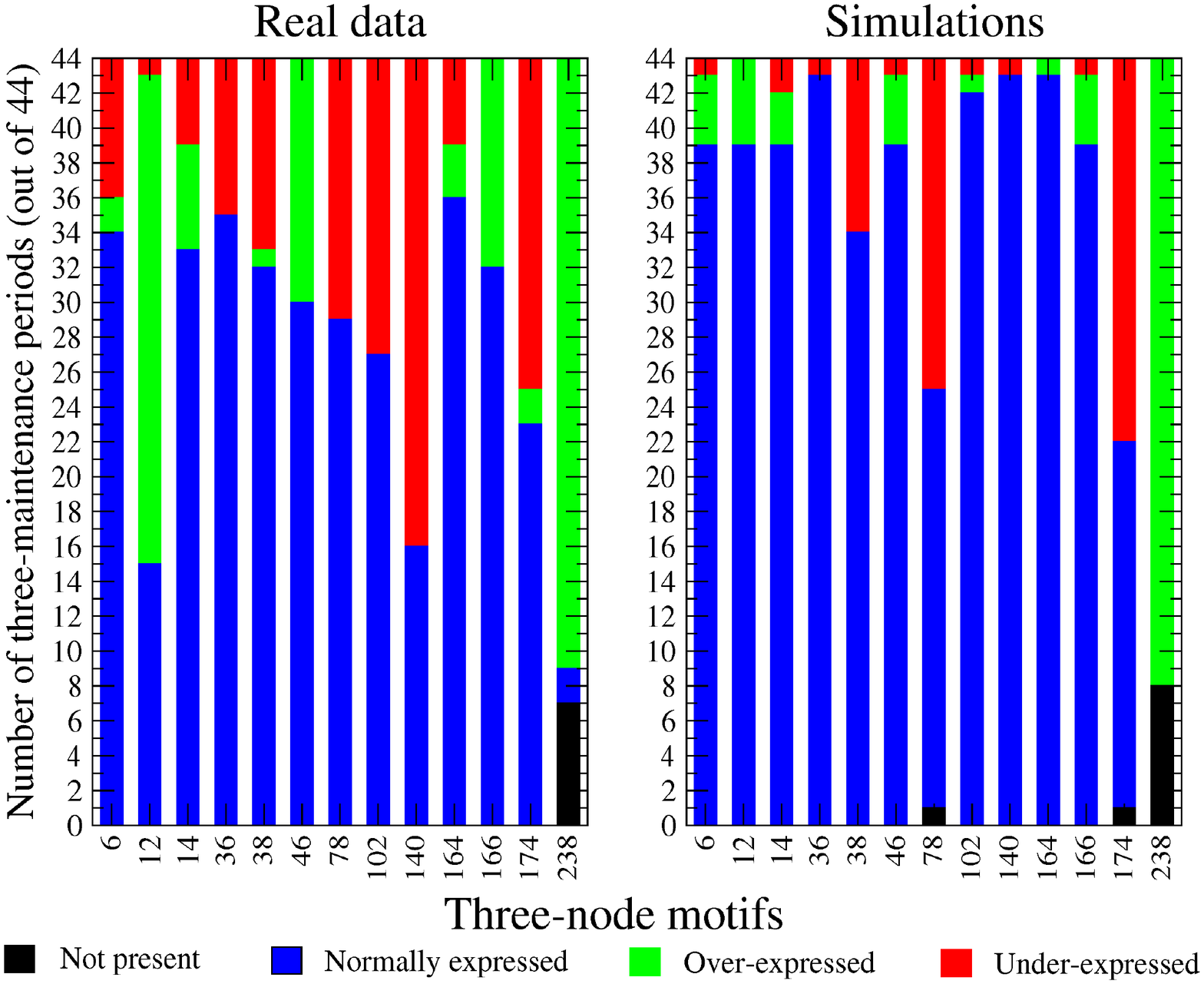}
\includegraphics[scale=0.36]{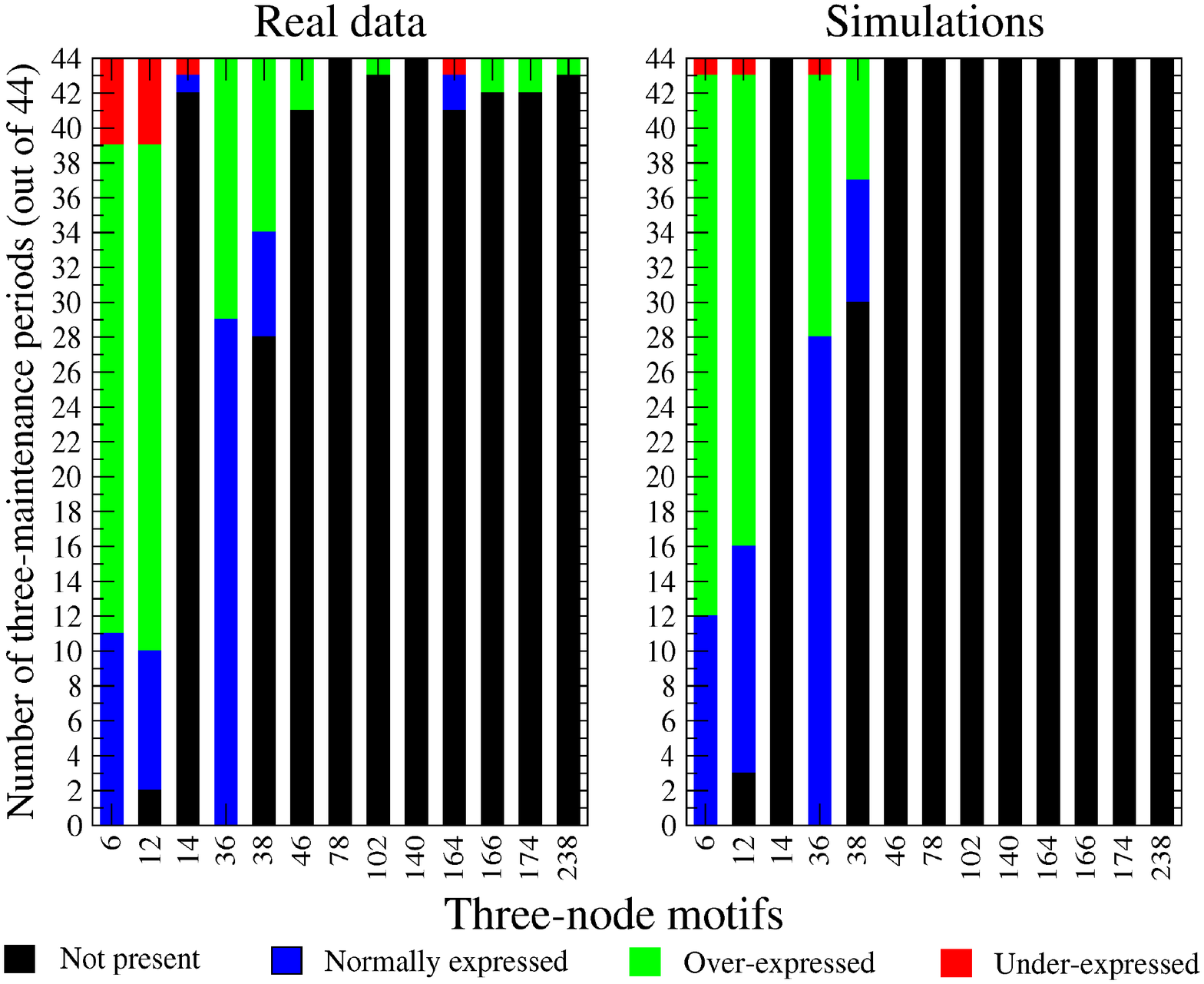}
}
                       \caption{Number of three-maintenance periods (out of 44) in which each three-motif type (indicated in the horizontal axis) is over-expressed, under-expressed, normally expressed, and not present in the original (top left panel) and Bonferroni networks (bottom left panel) associated with real data of \emph{borrower-aggressor}  and with \emph{borrower-aggressor} transactions from simulations of the model with $w=1$ (original (top left) and Bonferroni (bottom left)). Over-expressions and under expressions are obtained by performing a multiple hypothesis test correction. The over/under expression of a three motif indicates that the corresponding $p$-value provided by FANMOD was smaller than $0.01/(13 \cdot 44)$, where $13$ is the number of three-motif types and $44$ is the number of three-maintenance periods investigated.}  \label{threemotifsborr} 
\end{center}
\end{figure}

Fig.\ref{threemotifslend} shows the number of three-maintenance periods in which each 3-motif type turns out to be over-expressed, under-expressed, normally expressed, and not present in the original (top panels) and Bonferroni (bottom panels) networks obtained from real data of lender-aggressor overnight transactions among Italian banks (left panels) and from lender-aggressor transactions of a realization of our model with $w=1$. A comparison of top panels of Fig. \ref{threemotifslend} indicates that the number of three maintenance periods in which each motif is normally expressed in the original network obtained from simulations is larger than the corresponding number obtained from real data. The only exception is motif 174. This motif is normally expressed in one three-maintenance period in the original network from simulations, while it is normally expressed in four three-maintenance periods in the original network from real data. It involves two bidirectional links plus one directional link that closes the triangle. The reason why it is under-expressed in most of three-maintenance periods in the original network of both real data and simulations may be that the lack of reciprocity in the third connection is disfavored with respect to full reciprocity, which is accounted by motif 238. This motif is indeed over-expressed in most three-maintenance periods in both real data and simulations. Taking into account the fact that, on average, the frequency of motifs 174 ($< 1\%$) and 238 ($< 0.1\%$) is small in both simulations and real data, results observed for these two 3-motifs may be the result of the competition of two interdependent factors. The first one is that a $p$-value is automatically associated with a motif by the tool FANMOD, by comparing the frequency of the given motif in the network with its frequency in 1000 networks obtained by randomly rewiring the original one, without taking into account the 
link weight, that is, the number of transactions associated with a directed link between two banks. The second factor is the heterogeneity of banks, which makes it likely that two large banks interact with others just because they trade a lot, both as a borrower and as a lender. This result is also supported by the results reported in the bottom panels of Fig. \ref{threemotifslend}, where 3-motifs in the Bonferroni network of over-expressed trading is concerned. Indeed, motifs 174 and 238 never appear in the Bonferroni network from simulations, and they do not appear in, at least, 35 out of 44 three-maintenance periods in real data. It is also to mention that, for the motifs that appear in Bonferroni network, the number of times in which any three-motif is either over- or under-expressed in the networks obtained from simulations is rather small. On the contrary, in real data we notice that some 3-motifs labeled as  6, 12, 36, and 174 are over-expressed in at least 10 out of 44 three-maintenance periods.
 Such a difference between real data and simulations, which is observed in both the original networks and Bonferroni networks, suggests the presence of a non-trivial structure of nodes, e.g. communities, in real data, which is not captured in the model. A similar outcome is observed when we focus on borrower-aggressor transactions (see Fig.\ref{threemotifsborr}), though, in this case, the statistics is much smaller than in the case of lender-aggressor transactions, especially in the Bonferroni network, and, therefore, the impact of banks heterogeneity is stronger.
 
\section{A model with limited memory}\label{finitemem}

\begin{figure} [H]
\begin{center}
                      {\includegraphics[scale=0.36]{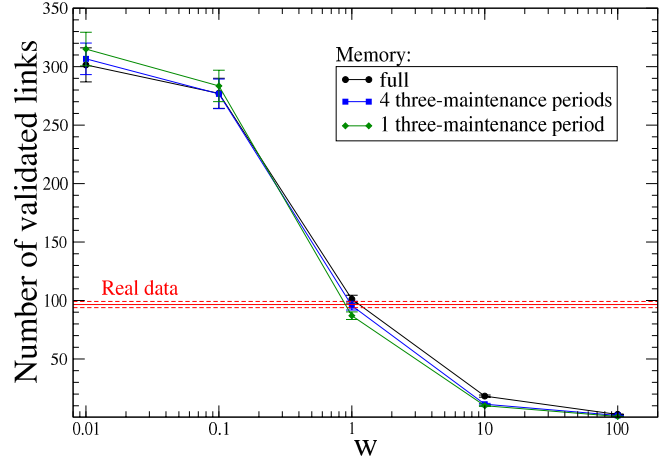}
                       \includegraphics[scale=0.36]{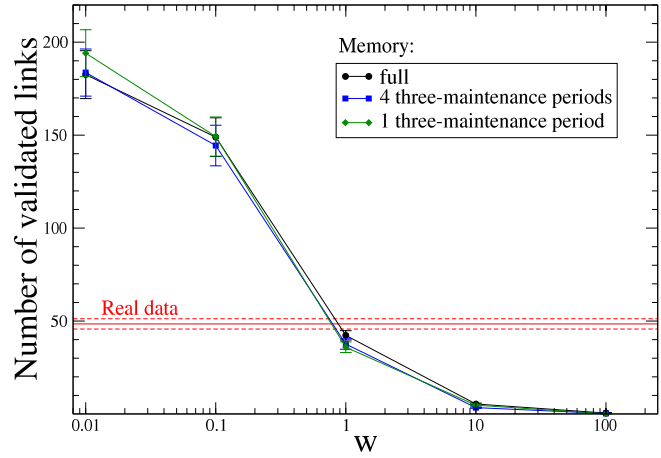}}
\caption{Average number of statistically validated links associated with simulations of transaction data across 44 three-maintenance periods at varying values of parameter $w$ for different settings of the finite memory parameter $Q$: memory of the \emph{full} set of past transactions, $Q=4$ three-maintenance periods in the past, and $Q=1$ three-maintenance period in the past. Error bars correspond to one standard deviation of the average. Top panel refers to lender-aggressor transactions, while bottom panel to borrower-aggressor transactions.The solid red line is the average value observed in real data of transactions among italian banks, and dashed lines, indicate the average plus and minus one standard deviation of the average.}  \label{averagelinksw} 
\end{center}
\end{figure}

The existence of a steady state in our model is a key aspect to tackle on. %
While a detailed analysis of the model's long-run results is still on the way, a qualitative discussion %
can already give some answers and suggests how to modify our assumption about the control parameters. %
When the heterogeneity of banks' transactions is constant over time, it is quite straightforward to figure out %
the evolution of the system. Let us consider a system consisting of just one bank $A$ which acts as borrower and a set %
of banks \{$\Phi$\}. Since the bank $B$ with higher number of transactions to execute in $\Phi$ is more likely to be selected %
as a partner from bank $A$, the parameter $N_{A \rightarrow B}(t)$ will start increasing. %
This process will go on and as $t \rightarrow \infty$ the system will move towards a deterministic steady state, where bank %
$A$ only transacts with bank $B$. The situation is different when the heterogeneity in the number of transactions %
is not constant, like in our empirical data, for example. In this case, even starting from the previous stationary state, %
we can imagine a shock of some nature hitting bank $B$, which then has to drastically reduce its number of transactions. %
In order to complete its own transactions, bank $A$ will have to fall back to another relationship which, given enough time %
can compete with the new one. Therefore, when dealing with heterogeneous and time-dependent number of transactions %
we should expect a long-run state characterized by a series of metastable states. 
However, the time scale for this switching %
can be extremely large when compared with reality, where banks have to react quickly to the fast dynamics of market state. %
Along this line, it also appears unrealistic to assume that all the events occurred a long time in the past influence the %
current behavior of a bank. For all these reasons, 
we are going to investigate how our model changes if we include a memory parameter to take into account %
only transactions between bank $A$ and bank $B$ in the past $Q$ time-windows. Specifically, we consider two values of parameter $Q$, one is $Q=4$ three-maintenance periods, which corresponds more or less to one year, and the other one is $Q=1$ three-maintenance period, which means a period of roughly three months. In Fig. \ref{averagelinksw} we compare the number of statistically validated links that result from simulations with different values of parameter $w$ and $Q$. The figure shows that the impact of finite memory on the number of validated links is marginal, at least with respect to the impact of parameter $w$. This result may be due to the competition of different factors. First of all, the heterogeneity of banks used in all the simulations, as measured by the number of transactions per three-maintenance period per bank, is the one observed in real data, and may affect in a similar way all the simulations through its changes across the 44 three-maintenance periods. A second factor may be the absence 
of external shocks in the simulations, which avoids abrupt changes in the behavior of banks, which turns out to be self-replicating at different time scales. A third factor concerns the size of finite memory that we have considered. It's possible that a time scale of 1 three-maintenance period is not sufficiently small to impair the full effectiveness of the memory mechanism.

\begin{figure}[H]
\centering
                \includegraphics[width=4.4cm]{Fig4a}
                \includegraphics[width=4.4cm]{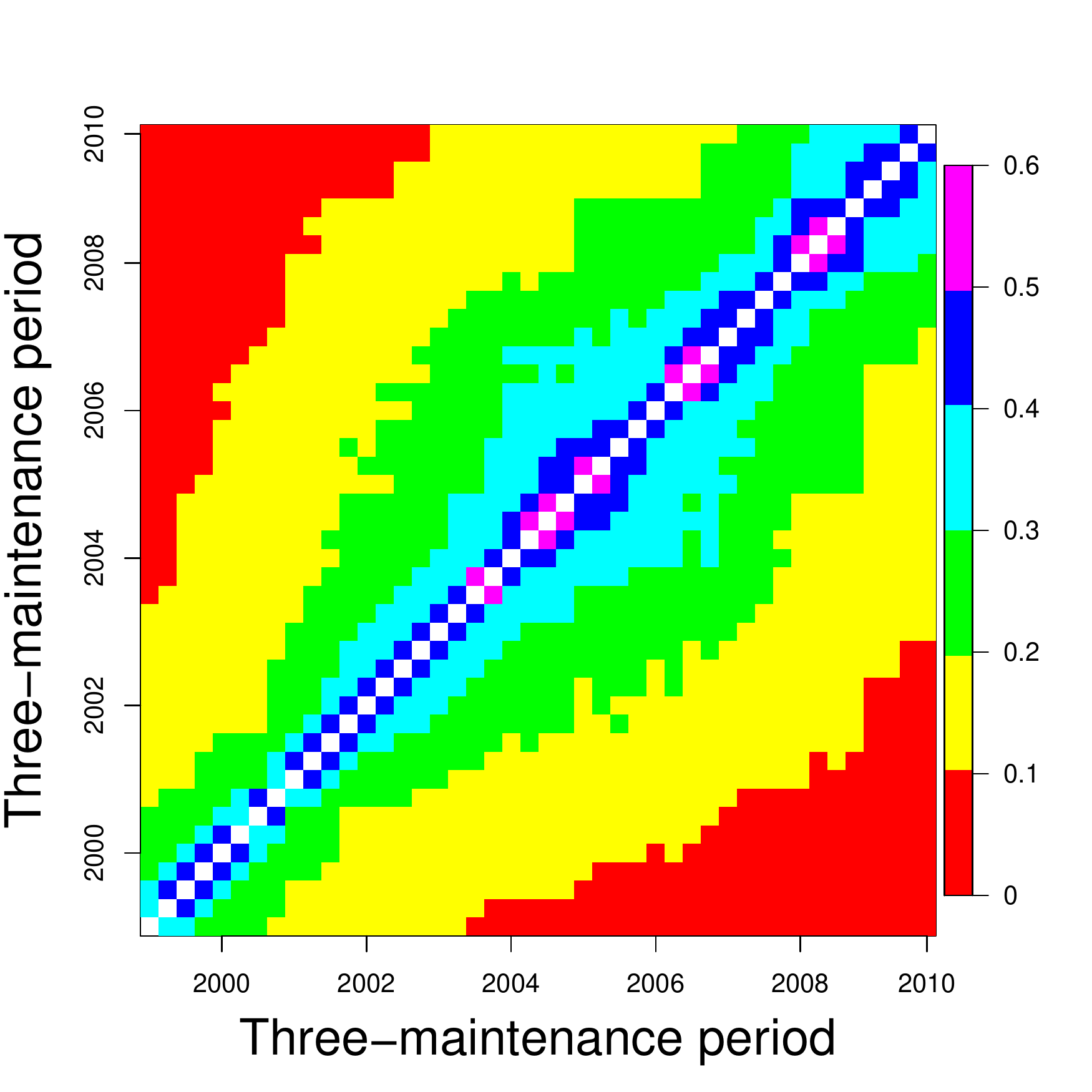}
                \includegraphics[width=4.4cm]{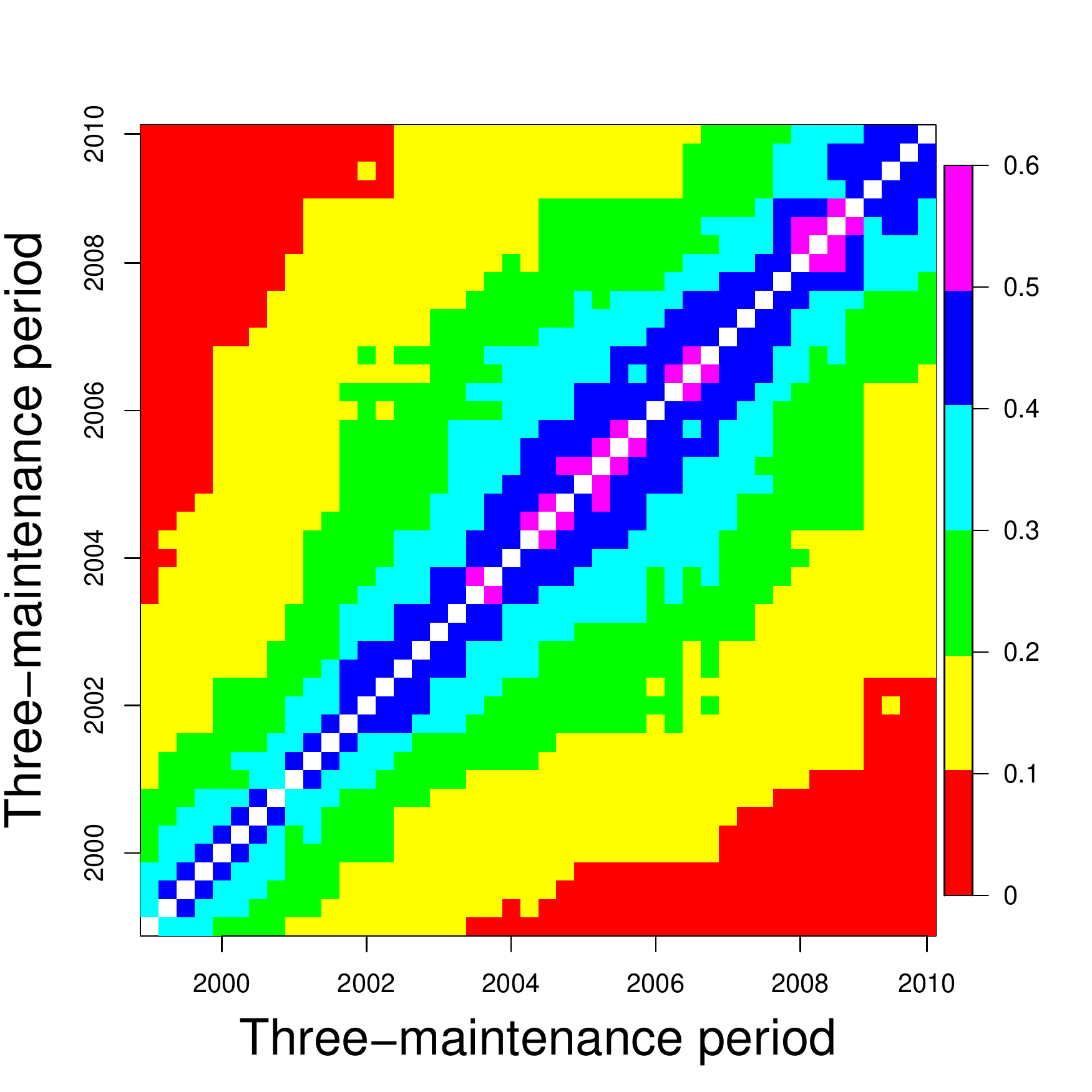}
                \caption{Lender-aggressor networks. Contour plots of the Jaccard index of links between all pairs of three-maintenance periods. The right panel shows the Jaccard index between original networks.  The central panel shows the Jaccard index between simulated networks with $Q=1$ and $w=1$. The right panel shows the Jaccard index between simulated networks with $Q=4$ and $w=1$.} \label{fig:M_contourmemfin}
\end{figure}

The presence of a memory impacts the degree of persistence of the simulated networks. In Fig. \ref{fig:M_contourmemfin} we show the contour plots of the Jaccard index of links between all pairs of three-maintenance periods. The right panel shows the Jaccard index between original networks.  The central panel shows the Jaccard index between simulated networks with $Q$=1 and $w$=1. The left panel shows the Jaccard index between simulated networks with $Q$=4 and $w$=1. From these plots it is quite evident that the persistence of links is not dramatically affected by the time window over which the level of memory extends in the past, although the case when $Q$=1 and $w$=1 shows overall lower levels of similarity of simulated networks, while the case when $Q$=4 and $w$=1 result is quite similar to empirical data. This would indicate that empirical data is better reproduced when the memory memory extends over the past, although it might not be necessary to have a full memory, as in the model of section \ref{model}.

\section{Conclusions}\label{conclusions}

We have introduced a simple model with memory to describe the formation of a networked structure of the e-MID market, which has been observed by Hatzopoulos et al. (2013). Such a structure presents preferential patterns of trading between banks incompatible with the null hypothesis of random pairing of banks. The null hypothesis takes into account the heterogeneity of banks, and so does the model. In fact, the model works at two different time scales, one at the level of single transactions between banks, and another one, at which the heterogeneity of banks  is set, according to their willingness to trade, as borrowers or lenders. The time scale considered in the paper is a three-maintenance period. The memory mechanism assumes that the probability that a lender and a borrower end up trading at a given time step depends on their overall willingness to trade (their heterogeneity) times the sum of two terms. The first one is the number of times in which the borrower 
borrowed from the lender in the past, and the other term, $w$, represents an overall attractiveness of borrowers. High values of parameter $w$ ($w>10$) disfavor the appearance of preferential patterns of trading. On the other hand, small values of $w$ ($w<0.1$) tend to freeze the market in a network in which most of the transactions occur between banks that have heavily traded in the past. So, low values of $w$ allow to model a status of the market in which ``trust'' (and ``distrust'') dominates the process of bank pairing. A high degree of agreement between model and real data, in terms of number of preferential links observed over time, is attained by setting $w=1$. It is interesting to note that, while the occurrence of preferential links is tested separately and independently for the set of borrower-aggressor and lender-aggressor transactions, the value $w=1$ that we used to simulate (at the same time) borrower 
aggressor and lender-aggressor transactions provides an excellent agreement with both sets of data. A comparison between the statistically validated networks obtained from simulations and real data, through the measure of the lagged Jaccard index among networks, indicates that, on average, the networked structure observed in a realization of the model is more persistent over the 44 simulated time windows than the networked structure observed in real data. This fact may be due to the relatively long time horizon of 11 years that we have considered, in which the e-MID market went through different phases, including a severe crisis, while parameter $w$ has been kept constant throughout the 44 simulated time windows.  The parameter $Q$ allows one to set the level of memory of banks. Three levels of $Q$ have been considered, $Q=1$ 
three-maintenance period, $Q=4$ three-maintenance periods and $Q= \infty$, which corresponds to the entire set of transactions since the beginning of a simulation. The comparison between model outcomes obtained with these values of $Q$ indicates that the model is not significantly affected by this parameter, at least in the investigated time horizon of 44 three-maintenance periods and in terms of the number of observed preferential connections. Model outcomes and real data have also been compared in terms of number of bidirectional links observed in the original and Bonferroni networks. The presence of bidirectional links is small ($<10\%$) in both real data and simulations, indicating a low degree of reciprocity in the system. However, on average, model outcomes present a smaller number of bidirectional links than real data. A simple method to introduce a tunable level of reciprocity has been proposed, but not investigated, in consideration of the low number of bidirectional links observed in real data. Finally, we have compared real data and simulations in terms of the presence of 3-motifs in the original and statistically validated networks. 3-motifs are simple structures that classify the 13 possible ways in which three nodes can be connected in a directed network. This analysis indicates that the presence of 3-motifs in the networks associated with model realizations is, on average, more similar to the one expected in a random network than the one observed in real data. This result suggests the moderate presence of a structure of connections in real data, such as clusters of banks, that is not captured by our model of random pairing with memory.
\appendix

\section{Details about Java Implementation}

 A version of the model  has been developed using Java and the \textbf{MASON} library %
 for multi-agent modeling (\cite{MASON}).
 The implementation within the Java/Mason framework will allow to include and integrate our model into the interbank sector %
 of the CRISIS macro-financial software library and in this appendix we briefly outline the structure of our Java/Mason software. %
 Besides the \emph{Model} and \emph{Scheduler} \textbf{MASON} classes, our model defines three new classes: %
 \textit{Transaction}, \textit{Bank} and  \textit{Market}. The class \textit{Transaction} basically represents a credit line between a \textit{Bank} which acts as a lender and %
 a \textit{Bank} which acts as a borrower. It contains data members to store the lending bank, the borrowing bank, the total 
 number of transactions occurred between these two banks since the beginning of the simulation (\textit{i.e.} %
 the memory parameter introduced in eq.(\ref{eq:la_lend_sel}), and also the number of transactions they had in the current %
 period of the simulation (identified with a three-maintenance period). Along with the appropriate getters and setters, 
 this class also provides a member function to increase by one both the marginal and cumulative transactions whenever a new deal %
 between  two banks occurs. %
 The class \textit{Bank} has data members to hold the number of transactions each bank wants to do as borrower %
 and as lender (two variables for the lender-aggressor scheme and two for the borrower-aggressor one) and four %
 ArrayLists to store the \textit{Transaction} objects describing the bank's credit lines with all the other banks. %
 It also contains methods to call when a new \textit{Transaction} occurs: they will update the cumulative and marginal %
 transactions between the two agents and correspondingly decrease the number of the transaction they can still perform %
 until the end of the simulated three-maintenance period. %
 The class \textit{Market} is an abstract class and is currently extended by two child classes: \textit{LA\_Market} %
 for the lender-aggressor setting and \textit{BA\_Market} for the case of borrower-aggressors. While these two classes %
 implement methods to choose the aggressor agent of the transaction and its counterpart following equations %
 \ref{eq:la_borr_sel} and \ref{eq:la_lend_sel} (or equations \ref{eq:ba_lend_sel} and \ref{eq:ba_borr_sel} for the borrower-aggressor scheme), %
 the parent abstract class defines a method to perform a single transaction, which automatically picks up a couple of agents %
 with the correct probability and updates all of their data members (cumulative and marginal transactions, and transactions to %
 be performed). %
 This greatly simplifies the code of the step() method in the \textit{Scheduler} class, which essentially consists in iterative %
 calls to the single transaction method until the number of transactions for both lender-aggressor and borrower-aggressor %
 scheme is exhausted. %

\section{Rewiring}

\begin{figure}[H]
\centering
                 {\includegraphics[width=4.0cm]{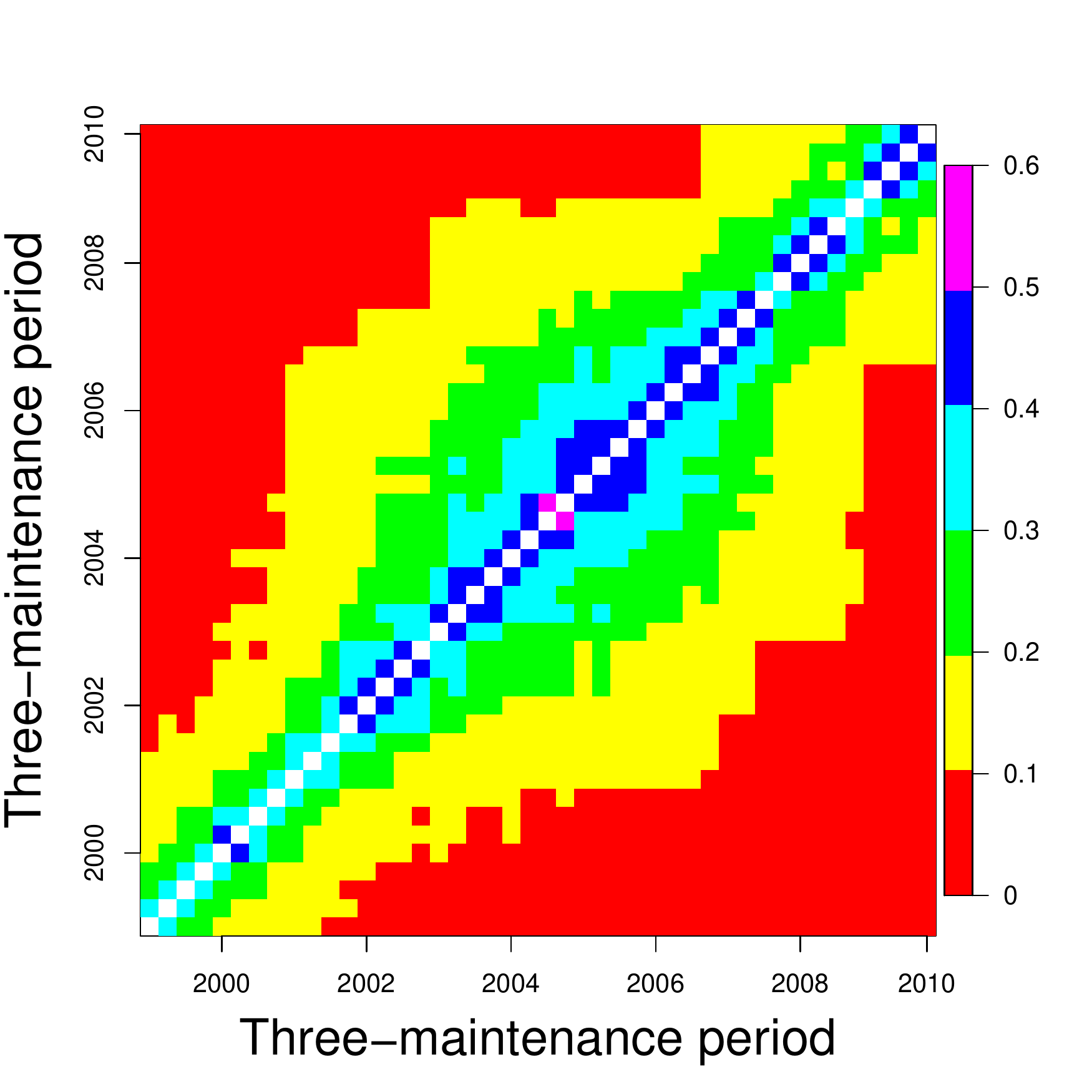}
                  \includegraphics[width=4.0cm]{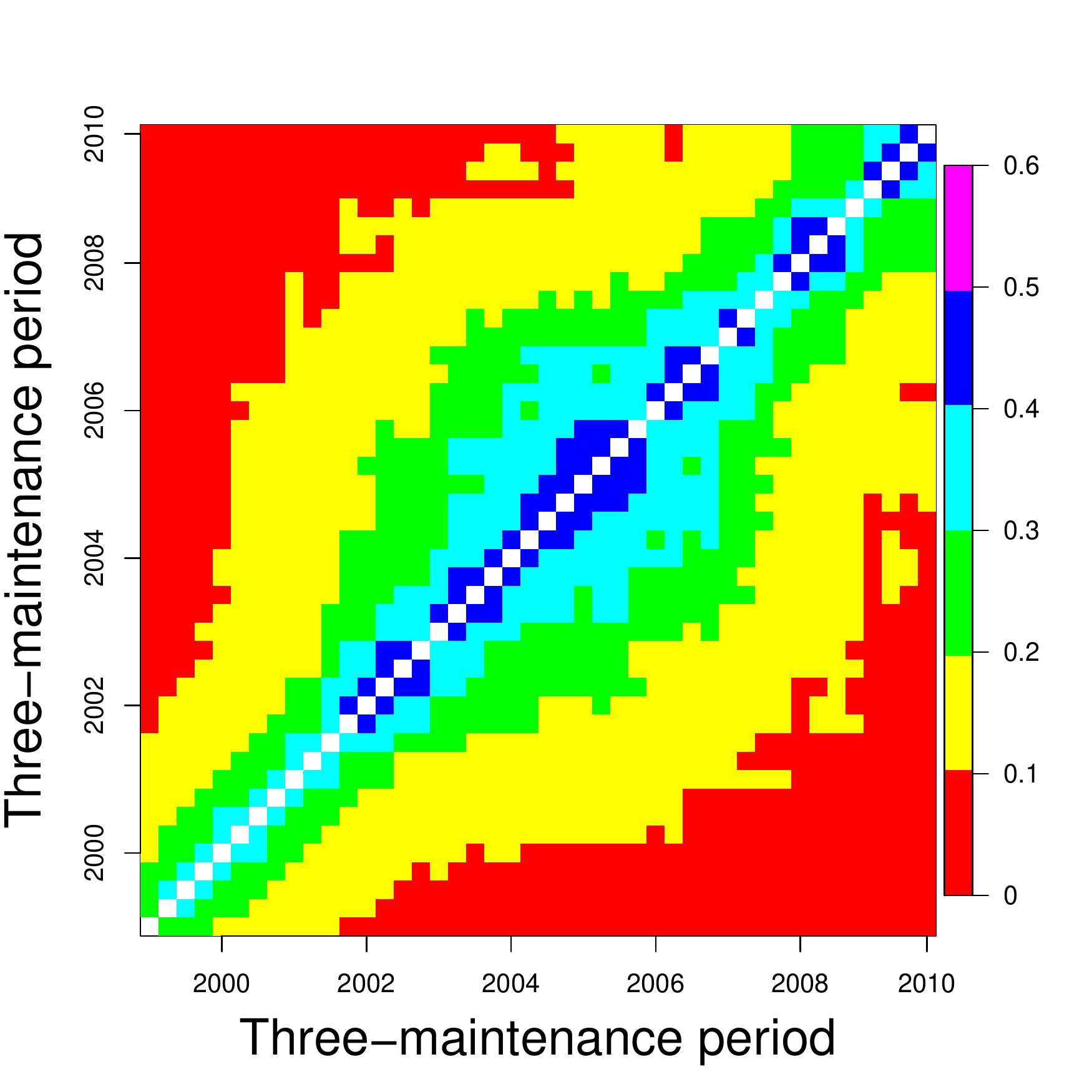} }\\
                 {\includegraphics[width=4.0cm]{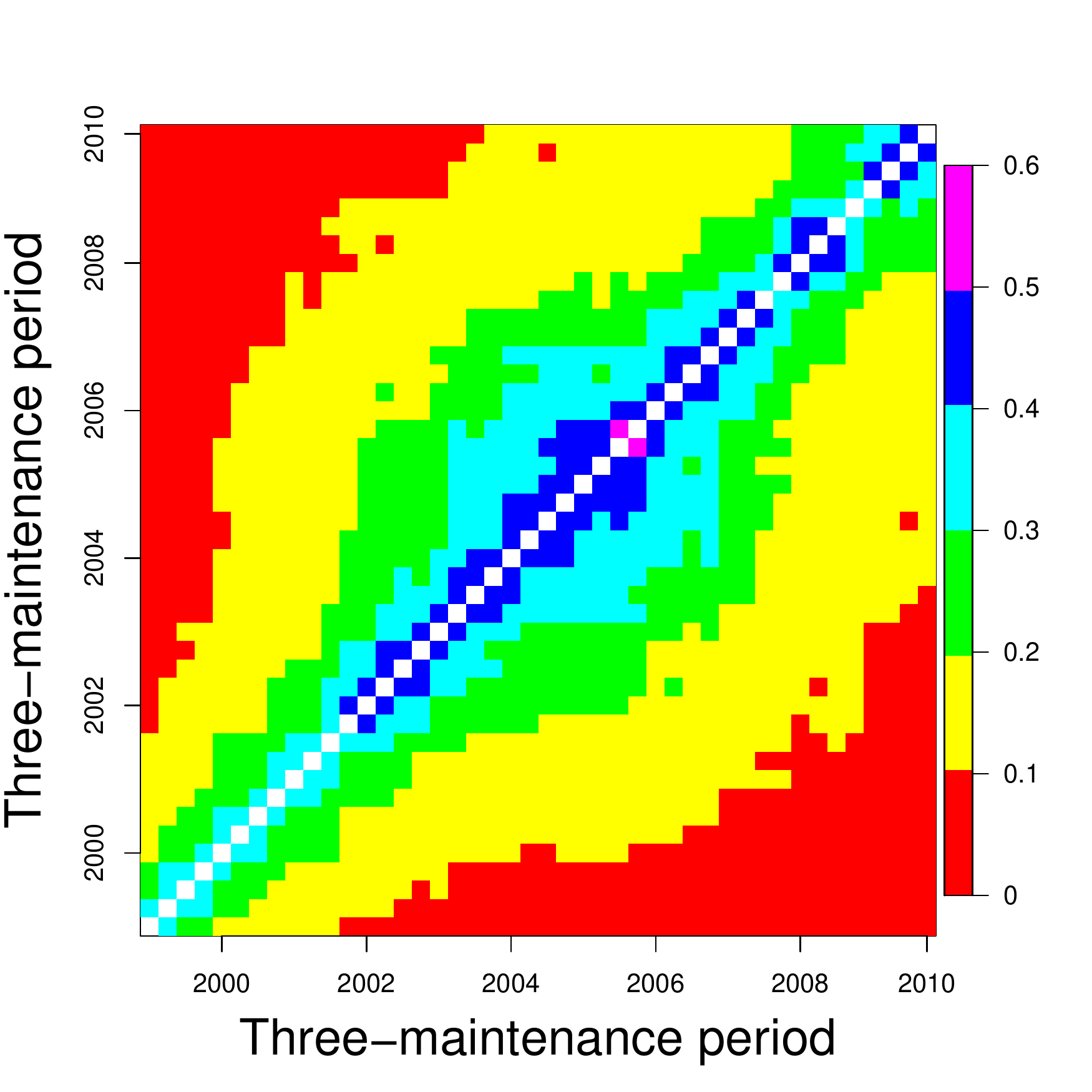}
                  \includegraphics[width=4.0cm]{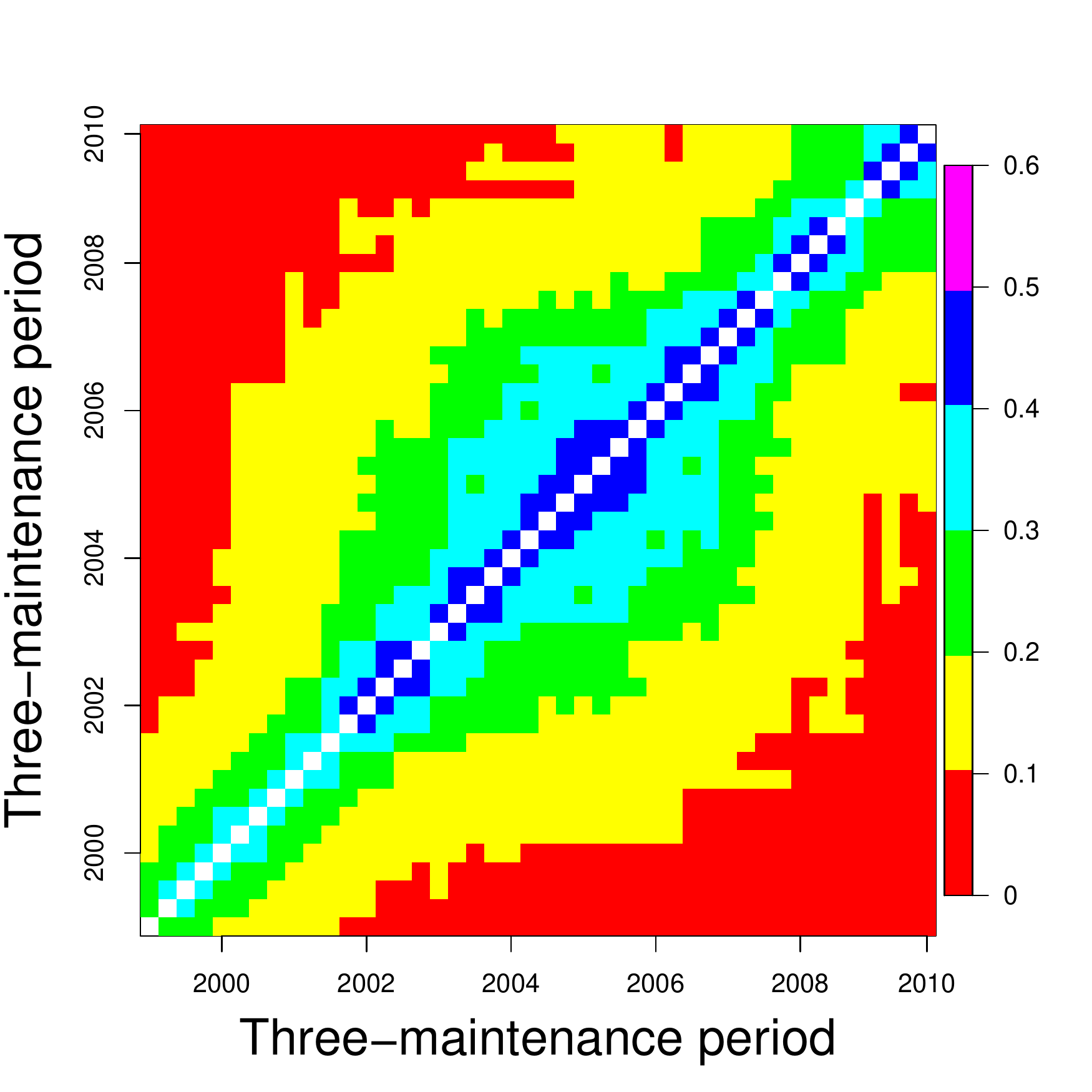} }\\
                 {\includegraphics[width=4.0cm]{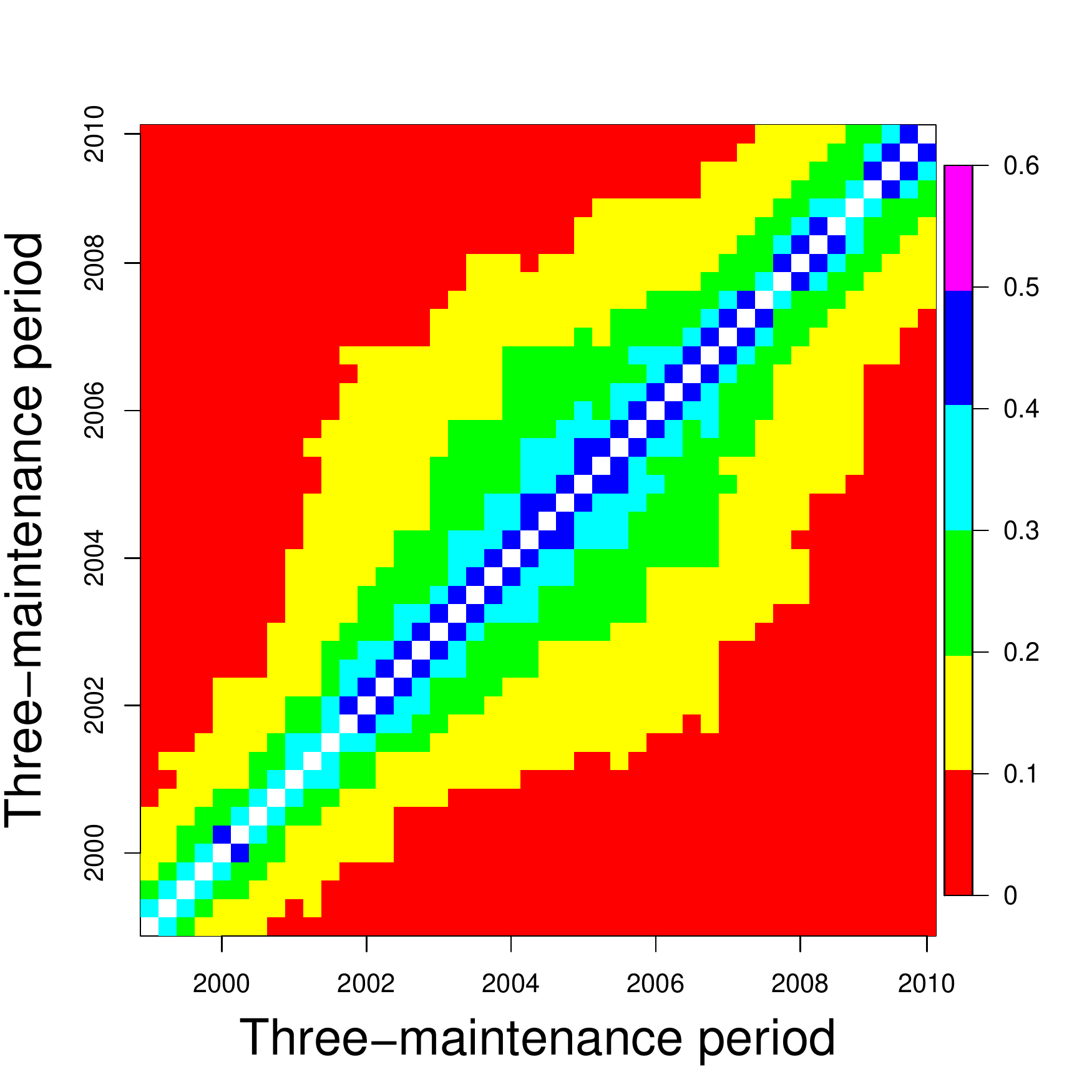}
                  \includegraphics[width=4.0cm]{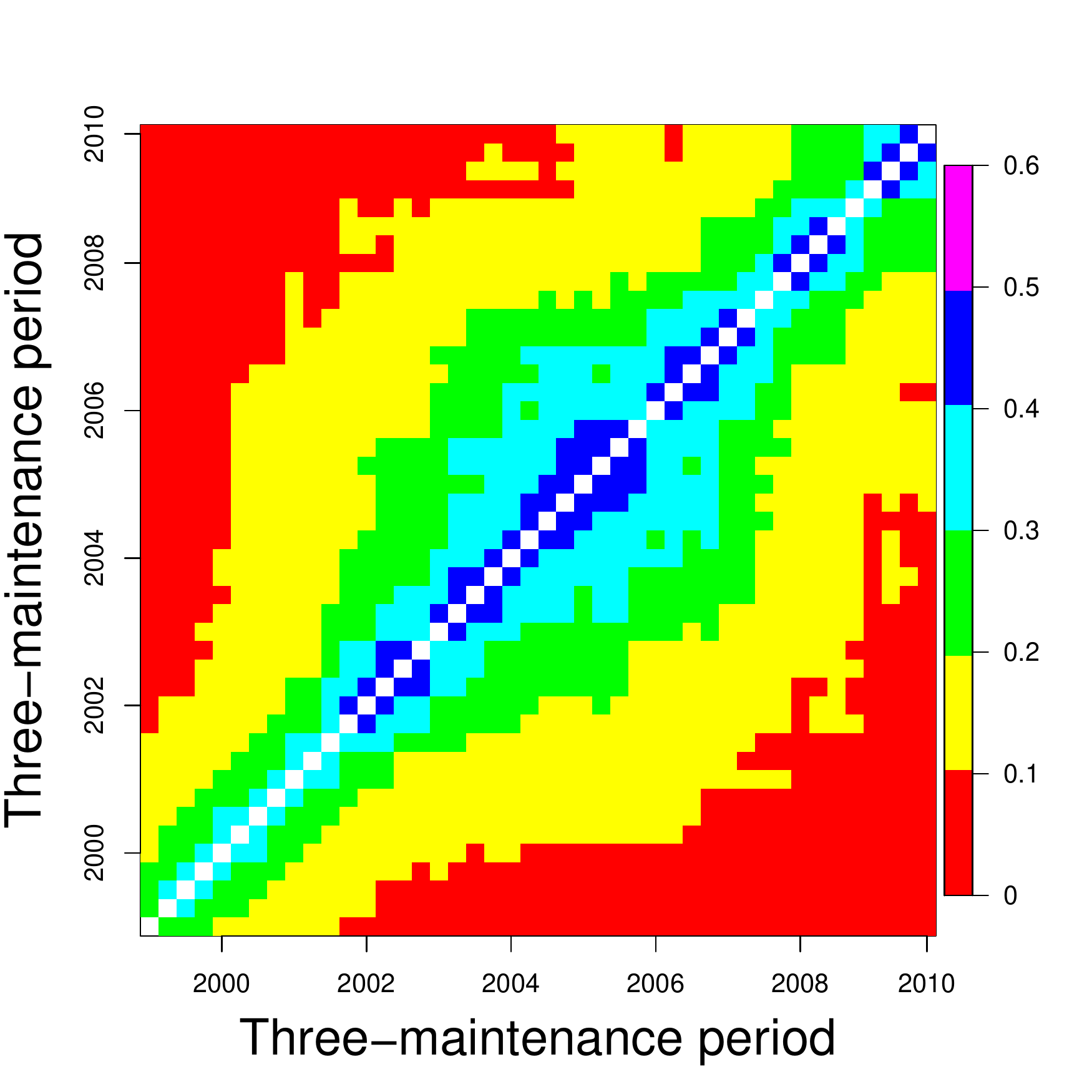} }\\
                 {\includegraphics[width=4.0cm]{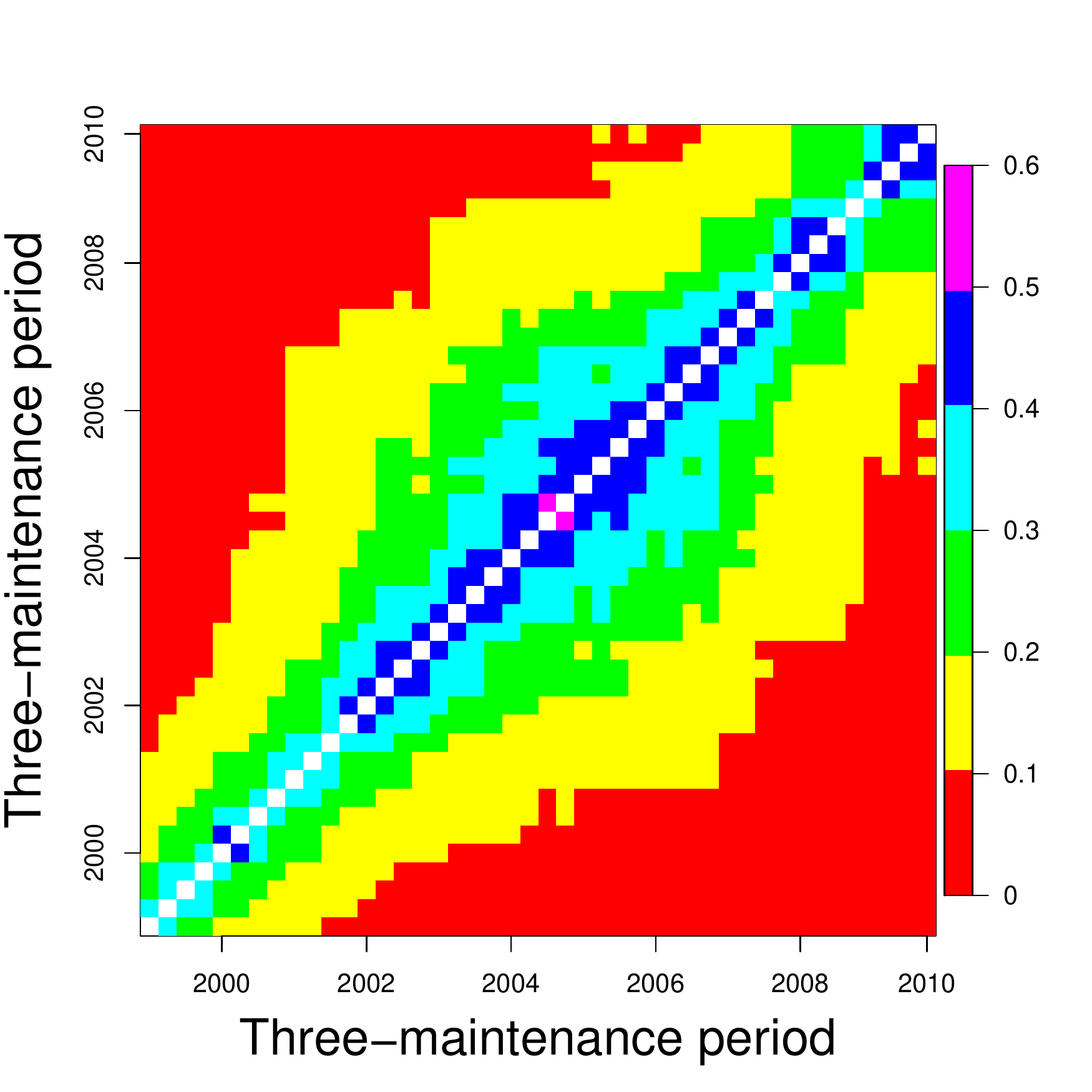}
                  \includegraphics[width=4.0cm]{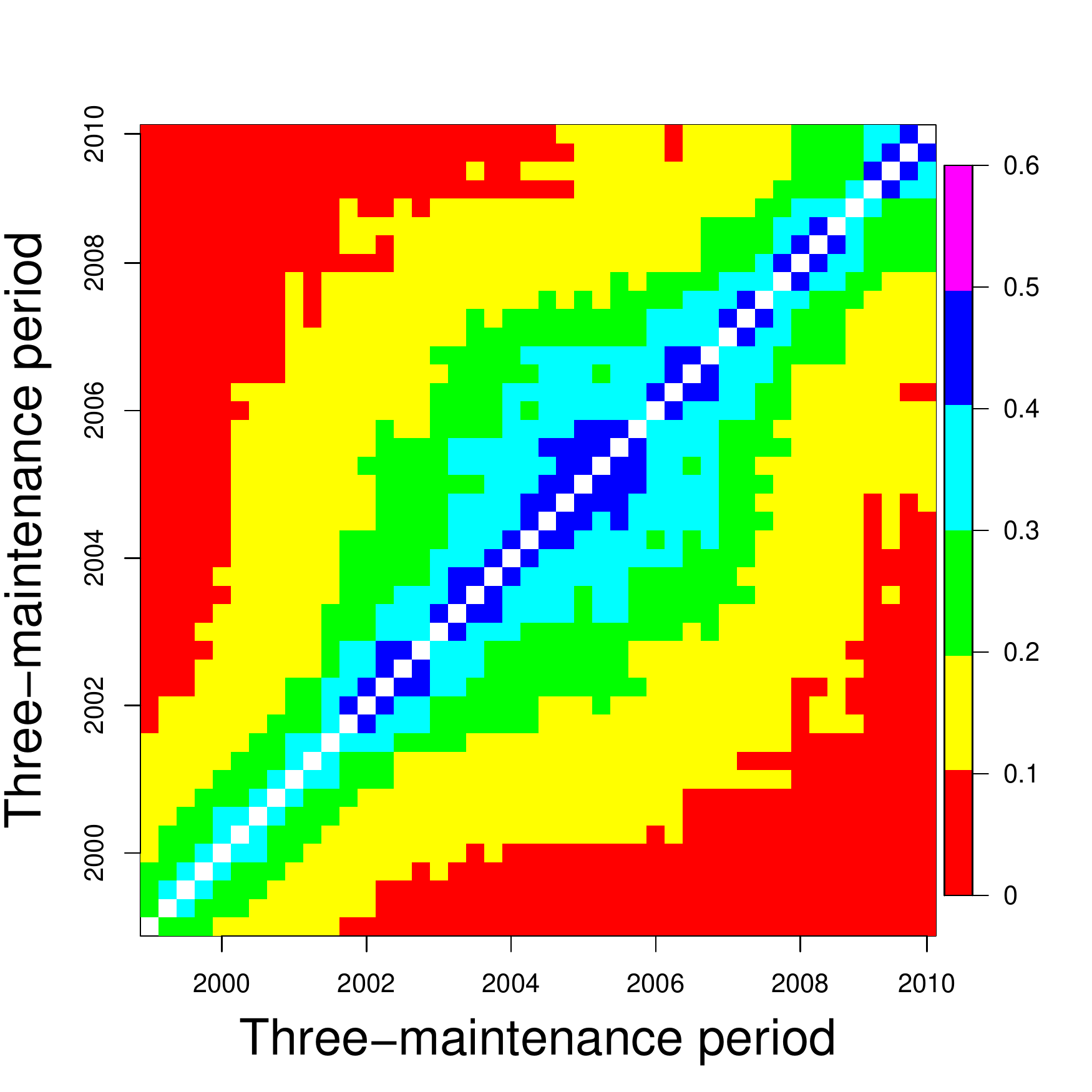} }
                \caption{Contour plots of the weighted Jaccard index of links between all pairs of three-maintenance periods for the lender-aggressor dataset. The top-left panel shows the weighted Jaccard index between original networks relative to the short-term maturities. The top-right panel shows the weighted Jaccard index of the corresponding rewired networks. The remaining left panels show the weighted Jaccard index original networks simulated by the model with full memory and $w=1$, $Q=1$ and $w=1$, $Q=4$ and $w=1$ from top to bottom. The right panels show the weighted Jaccard index of the corresponding rewired networks.} \label{fig:M_contourrew}
\end{figure}

The networks simulated by the model can be compared with those relative to empirical data by using the rewiring procedure illustrated in Ref. \cite{Hatzopoulos2013}. Such procedure consists in a re-shuffling of data that preserves the strength of each node. In Fig. \ref{fig:M_contourrew} we show the Contour plots of the weighted Jaccard index of links between all pairs of three-maintenance periods. 
The weighted Jaccard index has been introduced in Ref. \cite{Hatzopoulos2013} and generalizes the usual Jaccard index as to include the weight of each link. Here the weights are given by the number of transactions between a pair of banks. 
The top-left panel shows the Jaccard index between original networks. The top-right panel shows the weighted Jaccard index between re-wired networks. The remaining left panels show the weighted Jaccard index original networks simulated by the model with full memory and $w=1$, $Q=1$ and $w=1$, $Q=4$ and $w=1$ from top to bottom. The right panels show the weighted Jaccard index for the original simulated networks that have been subjected to a rewiring procedure.

By visual inspection one can notice that the set of parameters with $Q=4$ or with full memory are those that better reproduce the empirical data. Thus we confirm that empirical data are better reproduced when the memory memory extends over the past, although it might perhaps be not necessary to have a full memory as in the model of Section \ref{model}. 

\bibliographystyle{elsarticle-harv}




\end{document}